\documentclass[10pt,english,twocolumn,prl]{revtex4-1}
\usepackage[T1]{fontenc}
\usepackage[latin9]{inputenc}
\usepackage{mathrsfs}
\usepackage{amsmath}
\usepackage{amssymb}
\usepackage{graphicx}
\usepackage{xpatch}
\usepackage{hyperref}
\usepackage[leftcaption]{sidecap}
\hypersetup{colorlinks,breaklinks,urlcolor=[rgb]{0.18,0.203,0.565},linkcolor=[rgb]{0.18,0.203,0.565},citecolor=[rgb]{0.18,0.203,0.565}}

\newcommand{\sref}[2]{\hyperref[#1]{\ref*{#1}(#2)}}

\makeatletter
\newcommand{\sw}{\hat{S}}
\newcommand{\swd}{\hat{S}^\dagger}
\newcommand{\swds}{\hat{\tilde{S}}^\dagger}
\newcommand{\sws}{\hat{\tilde{S}}}

\makeatother

\usepackage{babel}
\begin{document}

\title{Quantum Optics of Spin Waves through ac Stark Modulation}

\author{Micha\l{} Parniak}
\email{michal.parniak@fuw.edu.pl}
\affiliation{Faculty of Physics, University of Warsaw, Pasteura 5, 02-093 Warsaw,
Poland}
\affiliation{Centre for Quantum Optical Technologies, Centre of New Technologies, University of Warsaw, Banacha 2c, 02-097 Warsaw, Poland}

\author{Mateusz Mazelanik}
\email{mateusz.mazelanik@fuw.edu.pl}
\affiliation{Faculty of Physics, University of Warsaw, Pasteura 5, 02-093 Warsaw,
Poland}
\affiliation{Centre for Quantum Optical Technologies, Centre of New Technologies, University of Warsaw, Banacha 2c, 02-097 Warsaw, Poland}

\author{Adam Leszczy\'{n}ski}
\affiliation{Faculty of Physics, University of Warsaw, Pasteura 5, 02-093 Warsaw,
Poland}
\affiliation{Centre for Quantum Optical Technologies, Centre of New Technologies, University of Warsaw, Banacha 2c, 02-097 Warsaw, Poland}

\author{Micha\l{} Lipka}
\affiliation{Faculty of Physics, University of Warsaw, Pasteura 5, 02-093 Warsaw,
Poland}
\affiliation{Centre for Quantum Optical Technologies, Centre of New Technologies, University of Warsaw, Banacha 2c, 02-097 Warsaw, Poland}

\author{Micha\l{} D\k{a}browski}
\affiliation{Faculty of Physics, University of Warsaw, Pasteura 5, 02-093 Warsaw,
Poland}
\affiliation{Centre for Quantum Optical Technologies, Centre of New Technologies, University of Warsaw, Banacha 2c, 02-097 Warsaw, Poland}

\author{Wojciech Wasilewski}

\affiliation{Faculty of Physics, University of Warsaw, Pasteura 5, 02-093 Warsaw,
Poland}
\affiliation{Centre for Quantum Optical Technologies, Centre of New Technologies, University of Warsaw, Banacha 2c, 02-097 Warsaw, Poland}

\begin{abstract}
We bring the set of linear quantum operations, important for many fundamental studies in photonic systems, to the material domain of collective excitations known as spin waves. Using the ac Stark effect we realize quantum operations on single excitations and demonstrate a spin-wave analogue of Hong-Ou-Mandel effect, realized via a beamsplitter implemented in the spin wave domain. Our scheme equips atomic-ensemble-based quantum repeaters with quantum information processing capability and can be readily brought to other physical systems, such as doped crystals or room-temperature atomic ensembles.
\end{abstract}

\maketitle

The Hong-Ou-Mandel (HOM) interference \cite{Hong1987} is an inherently quantum two-particle effect serving as an important test of both nonclassicality of the input state as well as proper operation of the beamsplitter. While nowadays it is easily achievable with photons, recent experiments demonstrated similar quantum-interferometric properties of atoms \cite{Kaufman2014,Lopes2015,Brandt2018}, phonons \cite{Toyoda2015,Hong2017}, plasmons \cite{Heeres2013,Fakonas2014} and photons but in elaborate hybrid systems \cite{Jin2013,Chen2015,Qian2016,Hong2017,Gao2018,Vural2018,Zopf2018}. This progress illuminates the perspective to combine linear operations, that have always been simple for photons, and nonlinear operations, that can be engineered in material systems. A quantum memory (QM) for light, where photons are stored in the form of collective atomic excitations is a good candidate for a bedrock to realize this proposal facilitating both fundamental studies and applications in quantum networks. Substantial challenges emerge, however, since photonic quantum networks need to extensively utilize multiplexing techniques, exploring photonic spatial and temporal  structure, to achieve high performance \cite{Lundeen2011,Chrapkiewicz2016,Reimer2016,Maring2017,Munro2010}. Multimode QMs \cite{Ding2013,Pu2017,Parniak2017} can become part of such networks, but a requirement of implementing complex linear operations on stored excitations arises.

In this Letter we harness these material quasi-particles \textendash{}
collective atomic excitations known as spin waves (SW). We demonstrate
that the spatial structure of SWs can be manipulated
via the off-resonant ac Stark (ACS) shift. Through SW diffraction (cf. Kapitza-Dirac effect \cite{Kapitza1933}) based
beamsplitter transformation, we realize the Hanbury Brown-Twiss
(HBT) type measurement in the SW domain \cite{Brown1956}, demonstrating precise control and
nonclassical statistics of atomic excitations. Finally, we observe
interference of two SWs \textendash{} an analogue of the HOM effect for photons. Thanks to the reversible
photon-SW mapping via the Duan-Lukin-Cirac-Zoller (DLCZ) protocol
\cite{Duan2001}, these techniques enable encoding states from
a high-dimensional Hilbert space into the spatial structure of SWs to facilitate not only new quantum communication schemes \cite{Fickler2016},
but also high data rate classical telecommunication \cite{Wang2012,Richardson2013}. A quantum repeater equipped with such coprocessing capability could perform error correction \cite{Jiang2007,Chen2007,Zhao2007,Muralidharan2014} or small-scale computation on transmitted quantum data. 
\begin{figure}
\includegraphics[width=1\columnwidth]{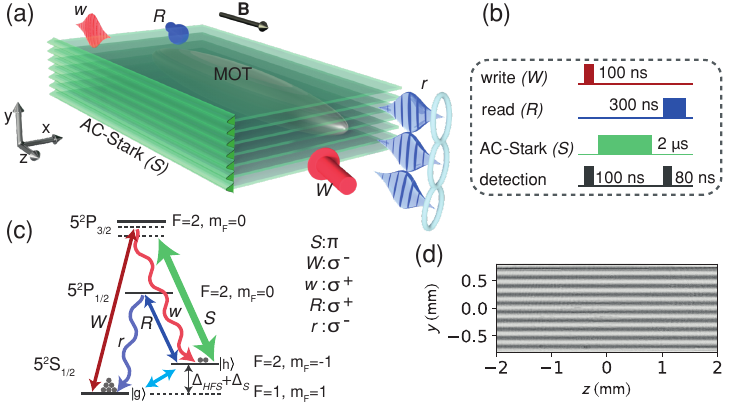}

\caption{Experimental setup for generating and manipulating
SWs: (a) detection of a single write-out photon $w$ scattered
from write laser $W$ heralds creation of a SW inside the atomic ensemble.
The SW is then manipulated using an ACS light pattern
(d) generated with a far-detuned laser $S$. The SW can then be converted by the read laser $R$ to a read-out photon $r$ with a reshaped spatial
mode; (c) the relevant energy level
configuration: $|g\rangle =|5^2S_{1/2}\ F=1, m_F=1\rangle$ and $|h\rangle=|5^2S_{1/2}\ F=2, m_F=-1\rangle$. The write laser is red-detuned from the $5^2S_{1/2}\ F=1\rightarrow 5^2P_{3/2}\ F=2$ transition by 25 MHz, the read laser is resonant with the $5^2S_{1/2}\ F=2\rightarrow 5^2P_{1/2}\ F=2$ transition and the ACS laser $S$ is red-detuned from the $5^2S_{1/2}\ F=2\rightarrow 5^2P_{3/2}$ line centroid by 1.43 GHz. During QM operation we keep constant bias magnetic field $\mathbf{B}=(50\ \mathrm{mG})\hat{e}_z$. (b) The sequence revealing the timing of each step during the experiment.\label{fig:trzyde}}
\end{figure}

The ability to perform beamsplitter transformations
with wavevector eigenmodes constitutes a full SW analogue of complex
linear-optical networks. The inherently nonclassical HOM interference with 80\% visibility is a concise demonstration of such transformation, which we realize with a three-way (in sense of the first three diffraction orders) splitter to demonstrate that SWs
always occupy either of the output modes. On the fundamental level the interference of two SWs with different wavevectors demonstrates preservation of coherence of many material quasi-particles in a thermal ensemble. Remarkably, the presented idea along with its applications can be brought to a multitude of physical systems where ACS shift control is feasible, including solids doped with rare-earth ions \cite{Bartholomew2018,Chaneliere2015}, color centers in diamond \cite{Acosta2013}, trapped ions \cite{Staanum2002} or warm atomic ensembles \cite{Moriyasu2009}.

\begin{figure}
\includegraphics[width=1\columnwidth]{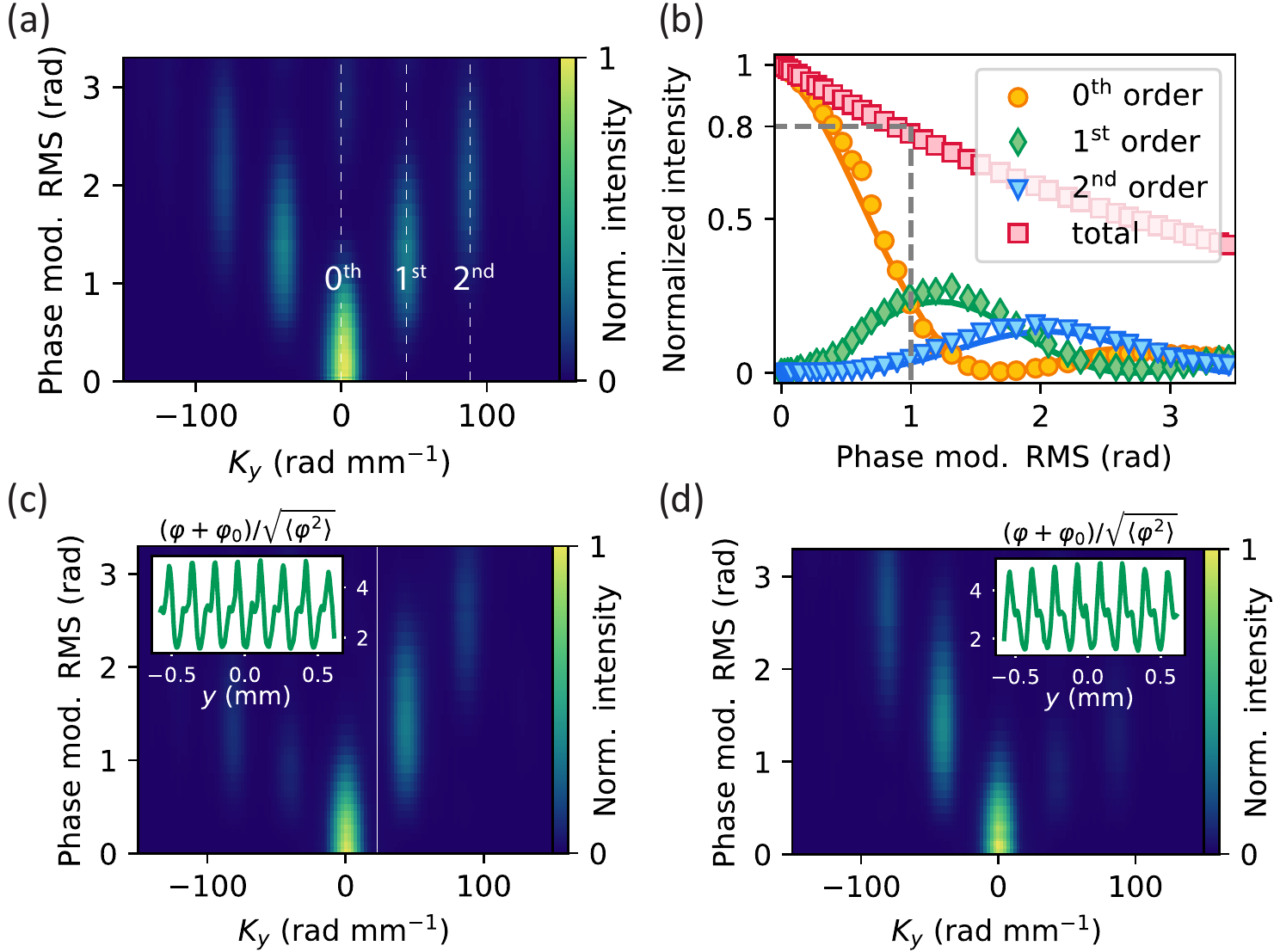}
\caption{Performance of the SW phase modulator: (a)
light intensity emitted from a SW as a function of a pure sine
modulation RMS amplitude $\sqrt{\langle\varphi_{S}^{2}\rangle}$ and
the wavevector $K_{y}$ component; (b) intensities in diffraction orders 0 to 2, marked in (a) along with the expected behavior (lines). In (c,d)
we change the modulation to include a term with higher frequency (insets - phase modulation patterns). Depending on the relative phase between the
two terms we
observe diffraction predominantly in the selected direction. \label{fig:modulator}}
\end{figure}

We use an elongated ($10\times0.3\times0.3\ \mathrm{mm}^3$) cold $^{87}$Rb ensemble to generate,
store and process ground-state SWs. Generation of single SWs relies on Raman scattering, which forms the basis
of the DLCZ protocol \cite{Duan2001}. A scattering
event, registered as a ``write-out'' ($w$) photon with a wavevector
$\mathbf{k}^{w}$, heralds creation of a single SW excitation with a wavevector
$\mathbf{K}=\mathbf{k}^{W}-\mathbf{k}^{w}$, where $\mathbf{k}^{W}$
is the write laser wavevector. The creation operator for a SW with wavevector $\mathbf{K}$
is $\swd_\mathbf{K}=N^{-1/2}\sum_{n}^{N}e^{i\mathbf{K}\cdot\mathbf{r}_{n}}|h_{n}\rangle\langle g_{n}|$
where $|0\rangle=|g_{1}\ldots g_{N}\rangle$ with
$N\approx10^{8}$ atoms in the ground state [Fig. \sref{fig:trzyde}{c}]. 
%The wavelike character of SWs
%emerges due to the spatial phase dependence of the atomic coherence, through which
 One atomic ensemble can accommodate many independent spatial
SW modes. 
For SW detection we use a read laser $R$ (wavevector $\mathbf{k}^{R}$) pulse that converts
the SW into a ``read-out'' ($r$) photon with wavevector
$\mathbf{k}^{r}=\mathbf{K}+\mathbf{k}^{R}$. 

To engineer ground-state SWs in our spatially-multimode QM we employ an off-resonant strong laser shaped with a spatial light modulator \cite{Leszczynski2018}
(Fig.~\ref{fig:trzyde}), inducing a spatially-dependent differential
ACS shift $\Delta_{S}(\mathbf{r})$ between levels $|g\rangle$
and $|h\rangle$, directly proportional to light intensity.
With negligible absorption and a small
transverse size of the ensemble we assume a constant intensity
along the propagation axis $x$ of the $S$ beam and thus write
$\Delta_{S}(\mathbf{r})=\Delta_{S}(y,z)$. The ACS shift leads
the SWs to accumulate an additional, spatially-dependent phase
$\varphi_{S}(y,z)=\Delta_{S}(y,z)T$ over the interaction
time $T\sim 2\ \mu \mathrm{s}$ with a typical $\Delta_{S}/2\pi\sim36\ \mathrm{kHz}$ obtained with $35\ \mathrm{mW/cm^2}$ intensity of $S$ light detuned from the respective resonance by $\delta_{S}/2\pi=1.43\ \mathrm{GHz}$. Such a manipulation is equivalent to the following
transformation of the SW creation operator within the Heisenberg
picture: $\swds_\mathbf{K}  =N^{-1/2}\sum_{n}^{N}e^{i\mathbf{K}\cdot\mathbf{r}_{n}+i\varphi_{S}(\mathbf{r}_{n})}|h_{n}\rangle\langle g_{n}|
=  \int\mathscr{F}[e^{i\varphi_{S}(\mathbf{r})}](\mathbf{k})\swd_{\mathbf{K}+\mathbf{k}}\mathrm{d}\mathbf{k},$
where $\mathscr{F}$ represents the Fourier transform in the spatial
domain. With periodic $\varphi_S(\mathbf{r})$, the transformation
becomes a Fourier series, realizing a multi-output
SW beamsplitter in two momentum-space dimensions.

We first select $\varphi_{S}$
to be a sine wave $\varphi_{S}(y)=\chi\sin(k_{g}y+\vartheta)$, where
$k_{g}$ is the grating wavevector. For technical reasons the sine
modulation is accompanied by a constant component
$\varphi_{0}$. With such modulation all SWs are diffracted into subsequent orders with central $y$ wavevector
components $K_{y}+mk_{g},\ m\in\mathbb{Z}$ and amplitudes
of subsequent orders depending on strength of phase modulation
quantified by its root mean square (RMS) amplitude $\sqrt{\langle\varphi_{S}^{2}\rangle}$. For benchmarking
we generate a coherent SW state with excitation number $\bar{{n}}\approx10^{5}$,
by seeding the Raman process with a coherent state of light tuned to $|g\rangle\leftrightarrow |h\rangle$ two-photon transition along with the $W$ laser. In Fig. \sref{fig:modulator}{a} we depict wavevector-resolved intensity of
light emitted from the SWs as a function of phase modulation
strength. By integrating the intensities in the discernible diffraction
orders we compare the experimental result with the expected behavior
[Fig. \sref{fig:modulator}{b}], finding excellent agreement and confirming the proposed
mechanism for SW diffraction. 

For the purpose of quantum engineering of SWs, we
now show that through precise control of the phase modulation pattern
we achieve desired amplitudes of diffraction orders, creating
a controllable $1\mathrm{-to-}N$ quantum network, where the zeroth order remains one of the output ports. Figures \sref{fig:modulator}{c,d} depict
the wavevector-resolved SW density. With
this we show that SWs is predominantly diffracted in the
selected direction through a proper asymmetrical modulation, here composed of sine wave with two frequencies with controlled relative phase.

\begin{figure}[t]
\includegraphics[width=1\columnwidth]{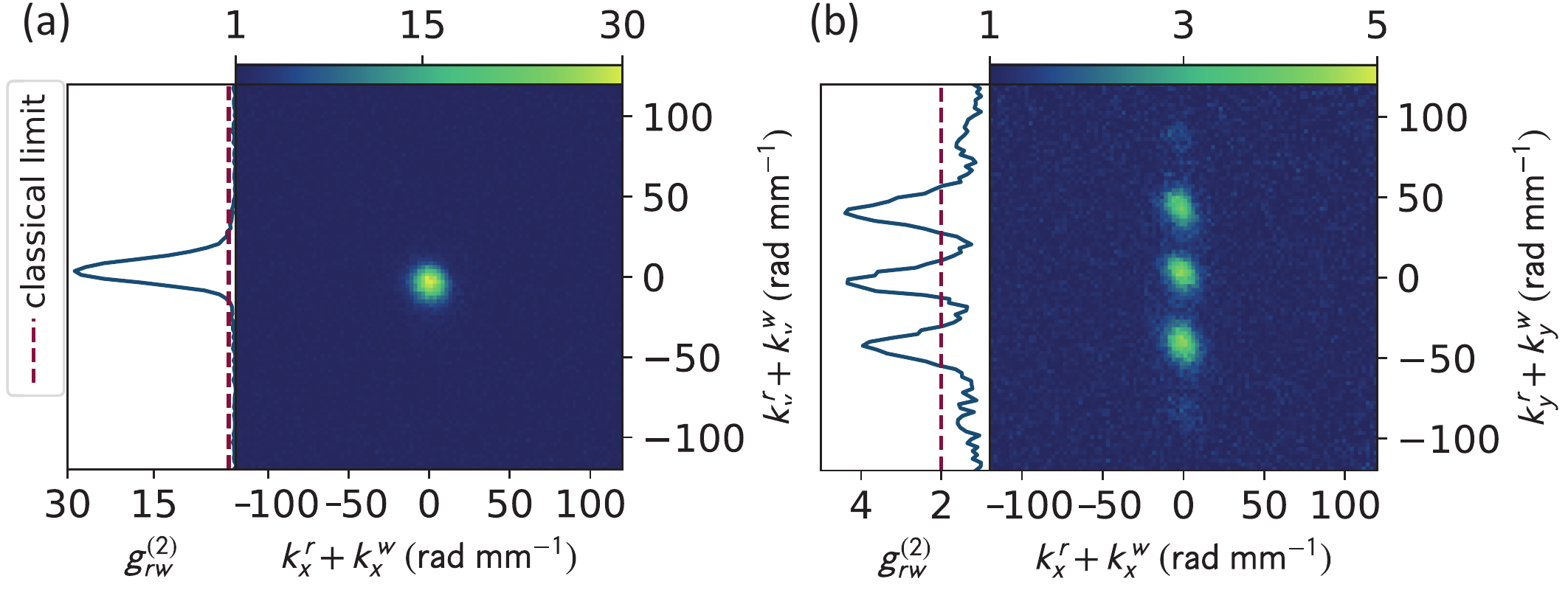}
\caption{A reference measurement (a) of second-order
cross-correlation $g_{rw}^{(2)}$ reveals a single peak at $k_{y}^{r}+k_{y}^{w}=0$,
demonstrating momentum anti-correlations.
By reshaping the SWs with a sine modulation pattern with
wavevector $k_{g}$, we modify the correlation function (b) to feature
two additional peaks at $k_{y}^{r}+k_{y}^{w}=\pm k_{g}$.\label{fig:giedwamapa} }
\end{figure}

With the SW modulation operating with high populations,
we now evaluate its performance at the single excitation level. We
probabilistically generate SWs heralded by detection of $w$
photons on an I-sCMOS camera situated in the far-field of the atomic ensemble. Quantum character of excitations is certified
by the second-order correlation function $g_{rw}^{(2)}=\langle\hat{n}_{r}\hat{n}_{w}\rangle/\langle\hat{n}_{r}\rangle\langle\hat{n}_{w}\rangle>2$,
which we express in terms of wavevector-sum variables, taking advantage
of wavevector multiplexing \cite{Parniak2017}. If the SWs are converted to photons without manipulation,
a single peak at $k_{x}^{r}+k_{x}^{w}=k_{y}^{r}+k_{y}^{w}=0$ is observed,
as in Fig. 3(a), since in general $\mathbf{k}^{w}+\mathbf{k}^{r}=\mathbf{k}^{W}+\mathbf{k}^{R}$
and we select $\mathbf{k}_{\perp}^{W}=-\mathbf{k}_{\perp}^{R}$. With sinusoidal phase modulation
with $\mathrm{RMS}=1.0\ \mathrm{rad}$ and wavevector $k_{g}$ applied along
the $y$-direction during storage, the peak is split into three equal
diffraction orders [Fig. \ref{fig:giedwamapa}(b)] with very little contribution to
higher orders, thus we may write that the $\sw_\mathbf{K}$ operator
is transformed into a sum of three operators: $\sws_\mathbf{K}=(\sw_\mathbf{K}+e^{i\vartheta}\sw_{\mathbf{K}+k_{g}\hat{e}_{y}}-e^{-i\vartheta}\sw_{\mathbf{K}-k_{g}\hat{e}_{y}})/\sqrt{3}$.
We certify quantum photon-number correlations in each
peak, demonstrating that our modulation scheme preserves statistical
properties of a SW, by operating with high efficiency and without
adding spurious noise.

We now use the presented manipulation to observe interference of SWs. Following Fig. \ref{fig:dolek}, using single-mode photon counting avalanche photodiodes (see \cite{Suppl}) we select a pair of Gaussian-shaped modes (mode field radius $\sigma=10.3\ \mathrm{rad}\ \mathrm{mm}^{-1}$) for the
$w$ photon ($wa$ and $wb$) corresponding to SW modes ($ra$ and $rb$) with
$K_{y}^{ra/rb}=\pm\Delta K_{y}/2=\pm45\ \mathrm{rad\ mm^{-1}}$ and
equal $K_{x}^{ra}=K_{x}^{rb}\approx200\ \mathrm{rad\ mm^{-1}}$ ($\Delta K_{x}=0$). By heralding a pair
of $w$ photons, we generate a SW pair $\swd_{ra}\swd_{rb}|0\rangle=|11\rangle_{ra,rb}$.
With a proper phase modulation each SW gets equally distributed into three equidistant
modes. We select the grating period $k_{g}=\Delta K_{y}=90\ \mathrm{rad\ mm^{-1}}$,
so that after manipulation we may write operators for resulting modes
$rc$ and $rd$ as $\swd_{rc}=(\swd_{ra}+e^{-i\vartheta}\swd_{rb}-e^{i\vartheta}\swd_{va})/\sqrt{{3}}$
and $\swd_{rd}=(\swd_{rb}+e^{-i\vartheta}\swd_{vb}-e^{i\vartheta}\swd_{ra})/\sqrt{{3}}$
. Let us now assume that the modes are well-overlapped, that is $\Delta K_{x}=0$
and $\Delta K_{y}=k_{g}$ and modes $va$
and $vb$ with $K_{y}^{va/vb}=\pm\frac{3}{2}\Delta K_{y}$ reside in vacuum (no excitation is heralded in this modes and we neglect their thermal occupations). With the output state given by $\hat{\rho}_{rc,rd}=1/9|00\rangle_{rc,rd}\langle00|+2/9|01\rangle_{rc,rd}\langle01|+2/9|10\rangle_{rc,rd}\langle10|+4/9|\psi\rangle\langle\psi|$
with $|\psi\rangle=(e^{i\vartheta}|20\rangle+e^{-i\vartheta}|02\rangle)/\sqrt{{2}}$ the interference is observable
in the heralded cross-correlation $g_{rc,rd|wa,wb}^{(2)}=\langle\hat{n}_{rc}\hat{n}_{rd}\hat{n}_{wa}\hat{n}_{wb}\rangle\langle\hat{n}_{wa}\hat{n}_{wb}\rangle/\langle\hat{n}_{rc}\hat{n}_{wa}\hat{n}_{wb}\rangle\langle\hat{n}_{rd}\hat{n}_{wa}\hat{n}_{wb}\rangle$
counting coincidences between photons emitted from modes $rc$
and $rd$ \textendash{} these coincidences vanish due to quantum
interference. Simultaneously, the number of self coincidences quantified
by $g_{rc,rc|wa,wb}^{(2)}$ (or $g_{rd,rd|wa,wb}^{(2)}$) increases.

In the experiment we first set $g^{(2)}_{wr}\approx20$ and then apply the modulation
that yields all cross-correlations, such as $g^{(2)}_{wa,rc}=g^{(2)}_{wa,rd}\approx6$ [Fig. \sref{fig:dolek2}{c}]. The initial extrinsic readout efficiency, defined as the ratio of $w-r$ coincidences to the number of $w$ counts is 4\%, which corresponds to $\sim30\%$ intrinsic memory efficiency after correcting for losses and detection efficiency. The efficiency of the modulation (at all discernible orders) is over 80\%. With the initial coincidence $w-r$ rate of $40\ \mathrm{Hz}$ we detect from $0.1$ up to $0.5$ quadruple coincidences per minute. With our current optical depth of 200 (as measured at the closed $F=2\rightarrow F=3$ transition) we can achieve efficiencies of over 60\% for classical pulses, however since a large detuning and power of the $R$ laser with $\sim1\ \mu\mathrm{s}$ long pulses is required, we currently achieve better overall performance at 30\% efficiency with only $80\ \mathrm{ns}$ long pulses and the $R$ laser tuned on resonance, which is mainly due to dark counts as well as filtration performance. 

\begin{figure}[t]
\includegraphics[width=\columnwidth]{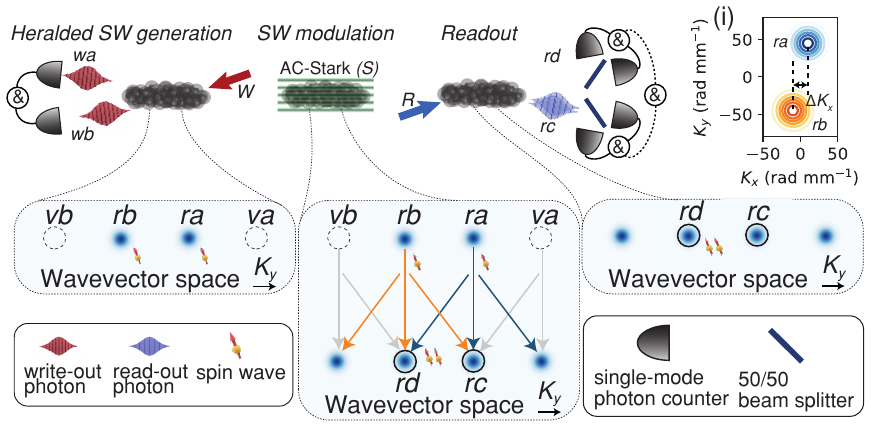}

\caption{The protocol for quantum interference of SWs.
Detection of two $w$ photons in modes $wa$ and $wb$ (selected through
single-mode fibers) heralds generation
of a SW pair in modes $ra$ and $rb$. The three-way splitter
is then used to interfere the two SWs. By detecting the SWs through photons converted to $rc$ and $rd$ modes we observe
bunching due to their bosonic nature. Inset (i) presents the input SW
modes in the ($K_{x}$, $K_{y}$) plane. 
Photonic detection modes
are always set to collect photons emitted from heralded SW
modes.\label{fig:dolek} }
\end{figure}

Figure \sref{fig:dolek2}{a} depicts the results obtained
as we change the overlap between shifted modes by varying $\Delta K_{x}$.
If the modes are overlapping at $\Delta K_{x}=0$,
we obtain a value of $g_{rc,rd|wa,wb}^{(2)}=0.20\pm0.06$, which
certifies the observation of two-SW HOM interference. Simultaneously, taking the $g_{rc,rc|wa,wb}^{(2)}$
auto-correlation we observe more than a two-fold increase from $0.5\pm0.4$
to $1.3\pm0.2$ compared with the case of non-overlapping modes, showing
that the pair of SWs is bunched and resides in a single mode. The theoretical prediction, detailed in~\cite{Suppl}, is made by first considering that each pair of contributing modes is squeezed to the same degree with the probability to generate a photon-SW pair $p=0.05$, then implementing the given beamsplitter network and finally adding the influence of dark counts at the detection stage.

A distinct quantum protocol is implemented by post-selecting only
$w$ photon detection events in the $wa$ mode [Fig.~\hyperref[fig:dolek2]{\ref*{fig:dolek2}(b)}]. With this, we effectively implement a
HBT measurement of a single SW in mode $ra$ without optical beamsplitting. The mode $rb$ is modeled as containing a thermal state $\hat{\rho}_{rb}(\bar{n})$ with $\bar{n}=0.1$.
Value of $g_{rc,rd|wa}^{(2)}=0.34\pm0.01<1$ clearly confirms the
single excitation character. As the modes are decoupled, we observe
a single photon statistics with $g_{rc,rc|wa}^{(2)}=0.67\pm0.08<1$
for the $rc$ mode and close to a single-mode thermal statistics with $g_{rd,rd|wa}^{(2)}=1.65\pm0.34$
for the $rd$ mode.

Finally, we directly populate the SW modes $ra$ and $rb$ with coherent state with population $\bar{n}=0.1$. The classical analogue of the HOM effect is observed [Fig.~\hyperref[fig:dolek2]{\ref*{fig:dolek2}(d)}] as we vary the phase offset of the ACS grating $\vartheta$ during a measurement, effectively creating a mixed state at the output. This corresponds to an interference of two phase-averaged coherent states, that yield an anti-correlated behavior $g^{(2)}_{rc,rd}\rightarrow0.5$ \cite{Jin2013}. In the experiment we indeed observe $g^{(2)}_{rc,rd}=0.53\pm0.02$ at $\Delta K_x=0$ which confirms the high visibility (47\% out of 50\% maximal). Note that a narrower distribution is observed in this case as we use a distinct mode function with $\sigma=6.8\ \mathrm{rad}\ \mathrm{mm}^{-1}$.

\begin{figure}
\includegraphics[width=\columnwidth]{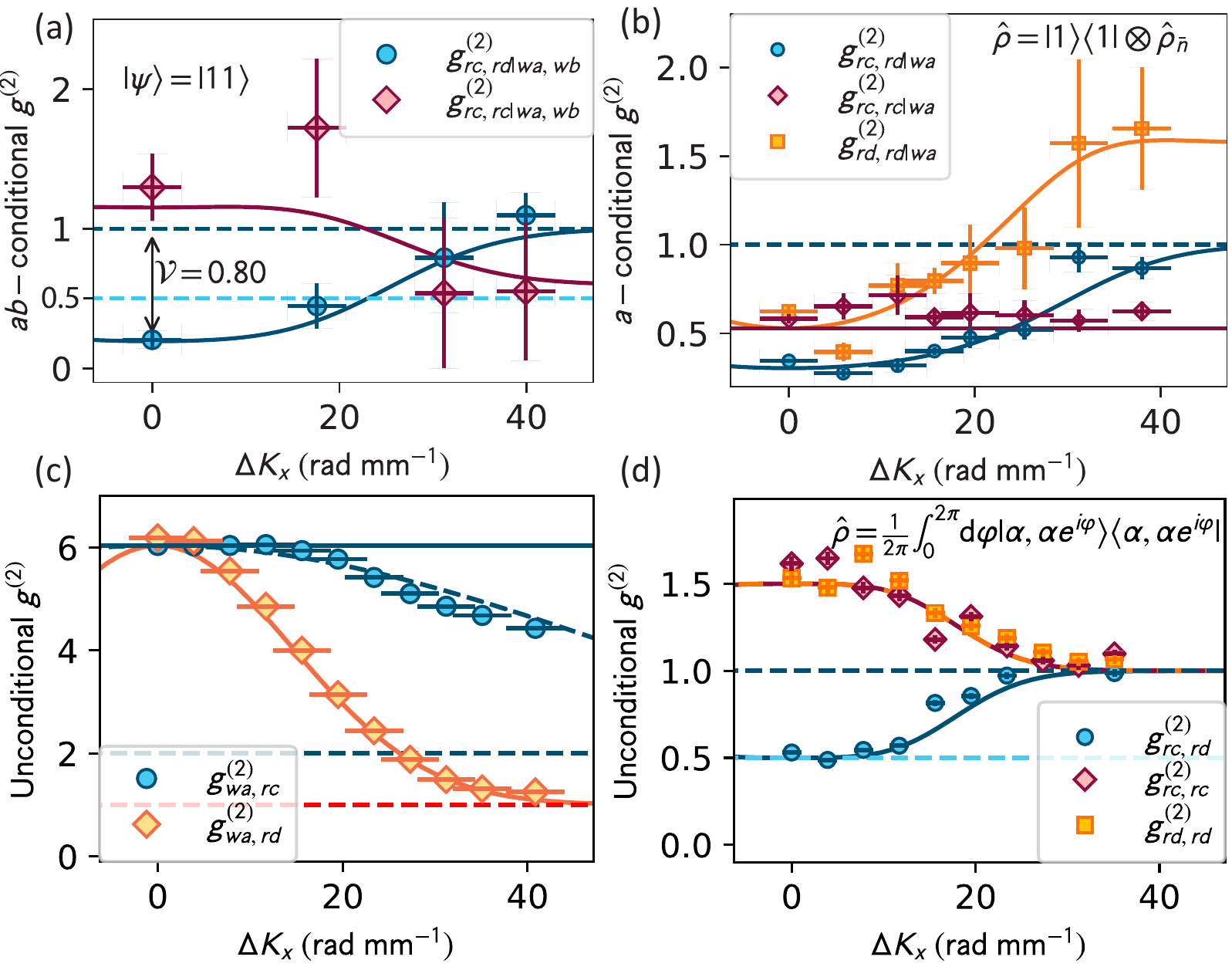}

\caption{Demonstration of quantum interference of SWs:
(a) HOM dip as a function of mode wavevector separation. Bunching may be suppressed if the modes $ra$ and $rb$ are
separated in the $K_{x}$ direction of the momentum space; (b) by heralding only the $w$ photon in the $wa$ mode, we implement a HBT experiment, observing nonclassical
statistics of the SW state; (c) the second-order correlation between $w$ and
$r$ photons validating the operation of the three-way
splitter with a slight drop in $g^{(2)}_{wa,rc}$ due to residual misalignment as the modes are moved (resulting in reduced fiber coupling efficiency); (d) HOM experiment for coherent input state with phase averaging. 
Vertical errorbars correspond to one standard deviation inferred from
Poissonian statistics of photon counts; horizontal errorbars
are due to mechanical precision of mode selection.\label{fig:dolek2}}
\end{figure}
This demonstration of HOM interference of SWs not only exposes
their bosonic nature, but paves the way towards implementing complex
quantum operations, including more SW modes, that are the primitives
of the linear-optical quantum computation scheme \cite{Ladd2010}.
The only hitherto successful attempt at HOM
interference of SWs relied on two different magnetic sublevels
coupled through Raman transitions \cite{Li2016}. Such approach could also be extended to the spatial domain, yet we believe that the ACS modulation provides more versatility in terms of implemented operations due to inherent access to all wavevectors.
Our experiment could furthermore greatly benefit from the deterministic SW generation protocol based on Rydberg blockade \cite{Li2016}, to improve our current heralded SW pair generation rate. Furthermore, an ultra-high OD or cavity-based design \cite{Cho2016,Bao2012,Hsiao2018} could bring the retrieval efficiency close to 90\%. The combination of the Gradient Echo Memory \cite{Hosseini2009,Hetet2015}
and the ACS modulation could enable using SWs
in the three-dimensional space, with their $K_{z}$ component coupled with the photonic temporal degree of freedom (pulses arriving at different times), and
transverse components of wavevectors paired with photonic
transverse coordinates.
 
With the proposed techniques a multiplexed source of heralded $l$-photon can be realized \cite{Nunn2013,Parniak2017} using a single atomic ensemble. A camera detector can herald 
creation of SWs in $l$ out of $M$ modes with wavevectors $(\mathbf{K}_1, \ldots,\mathbf{K}_l)$. Through adjustable ACS modulation we then realize a switch redirecting these $l$ populated modes into specific $l$ output modes, with the success probability given by the incomplete regularized Beta function $I_{p\eta_w}(l,M-l+1)\eta_r^l$ with $\eta_{w,r}$ being the efficiencies for detection of $r$ and $w$ photons. With $M \gtrsim l(1+3/\sqrt{l}) / p \eta_w$ the probability approaches $\eta_r^l$ and can dramatically beat the non-multiplexed scenario, even with source operating at much higher rates (see \cite{Suppl} for specific rate estimates). 

The same idea can be used to design a multiplexed quantum repeater following the proposals presented in \cite{Collins2007,Jiang2007,Chen2007,Zhao2007}. At the entanglement generation stage we combine the optical fields coming from two nodes at a beamsplitter and detect them with a camera. Detection of a $w$ photon with wavevector $k_w$ projects the pair of ensembles (A and B) into an entangled state $1/\sqrt{2}(\swd_{A,\mathbf{K}} + \swd_{B,\mathbf{K}})|0\rangle_A |0\rangle_B$, yet the generation rate can be high since we can keep a low probability to generate the state per mode, but achieve high rate per \textit{any} mode. In the similar manner as in the multiplexed photon generation protocol, many states can be generated at the same time. At the entanglement connection stage, the entangled states are paired to form entangled qubits and then ACS modulation matches the SW modes of two ensembles being connected so entanglement can be established between them via HOM interference of $r$ photons. Finally, with the ACS modulation we mix various entangled qubits stored in one ensemble to obtain purified pairs. This protocol corrects both for phase errors thanks to HOM interference and reduces errors in the logical space thanks to the entanglement purification step; yet most importantly it is inherently multiplexed and uses only a pair of atomic ensembles in each node (see \cite{Suppl} for more details).
More advanced error correction codes  for quantum repeaters have already been proposed \cite{Jiang2009,Munro2010,Glaudell2016,Muralidharan2014,Muralidharan2016}, but require a multi-qubit quantum computer at each node. Such advanced quantum information processing capability is hard to achieve in practice with linear optics \cite{Ladd2010}, but photons stored as Rydberg SWs for which nonlinear interactions can be engineered \cite{Peyronel2012a,Distante2016,Petrosyan2017} could provide such capability when combined with our linear-operations scheme.

Current parameters of the demonstrated device already allow realization of original schemes, but can be improved by better use of wavevector-multiplexing facilitating faster and nearly-deterministic generation of single SWs and prompt transit towards realization of the proposed protocols. In conclusion, our fundamentally new scheme of SW manipulation along with potential derivative protocols lends itself to many applications in light technologies and potentially allows exploration of nonlinear interactions in the spatial domain. 

\begin{acknowledgments}
\paragraph{Acknowledgments} We thank K.~Banaszek for generous support, M.~Jachura for proofreading
of the manuscript and K.~T.~Kaczmarek and P.~L.~Knight for helpful discussions. This work has been funded by the National Science
Centre, Poland (NCN) grants No. 2015/19/N/ST2/01671, 2016/21/B/ST2/02559,
2017/25/N/ST2/01163 and 2017/25/N/ST2/00713 and by the Polish Ministry of Science and Higher Education (MNiSW)
``Diamentowy Grant'' projects No. DI2013 011943 and DI2016 014846. M.D. was supported by the Foundation for Polish Science. The research is supported by the Foundation for Polish Science, co-financed by the European Union under the European Regional Development Fund, as a part of the ``Quantum Optical Technologies'' project carried out within the International Research Agendas programme.
\end{acknowledgments}
 M.P. and M.M. contributed equally to this work.

%\bibliographystyle{naturemag}
%\section*{References}
%\renewcommand\bibname{Methods References}

%\bibliographystyle{apsrev4-1}

%\renewcommand*{\bibfont}{\footnotesize}
%\footnotesize
\bibliography{bibliografia}

%%%%%%%%%% Merge with supplemental materials %%%%%%%%%%
\onecolumngrid
\clearpage
\vspace{0.8cm}
\begin{center}
	\textbf{\large Supplemental Material for ``Quantum Optics of Spin Waves through ac Stark Modulation''}
\end{center}
\vspace{0.8cm}
%%%%%%%%%% Merge with supplemental materials %%%%%%%%%%
%%%%%%%%%% Prefix a "S" to all equations, figures, tables and reset the counter %%%%%%%%%%
\setcounter{equation}{0}
\setcounter{figure}{0}
\setcounter{table}{0}
\makeatletter
\renewcommand{\theequation}{S\arabic{equation}}
\renewcommand{\thefigure}{S\arabic{figure}}

\renewcommand{\thesubsection}{S.\Roman{subsection}}
%%%%%%%%%% Prefix a "S" to all equations, figures, tables and reset the counter %%%%%%%%%%

\twocolumngrid
\subsection{Additional Discussion}
\paragraph{Directing photons into a specific mode}
\begin{figure}[h]
	\centering
	\includegraphics[width=1\columnwidth]{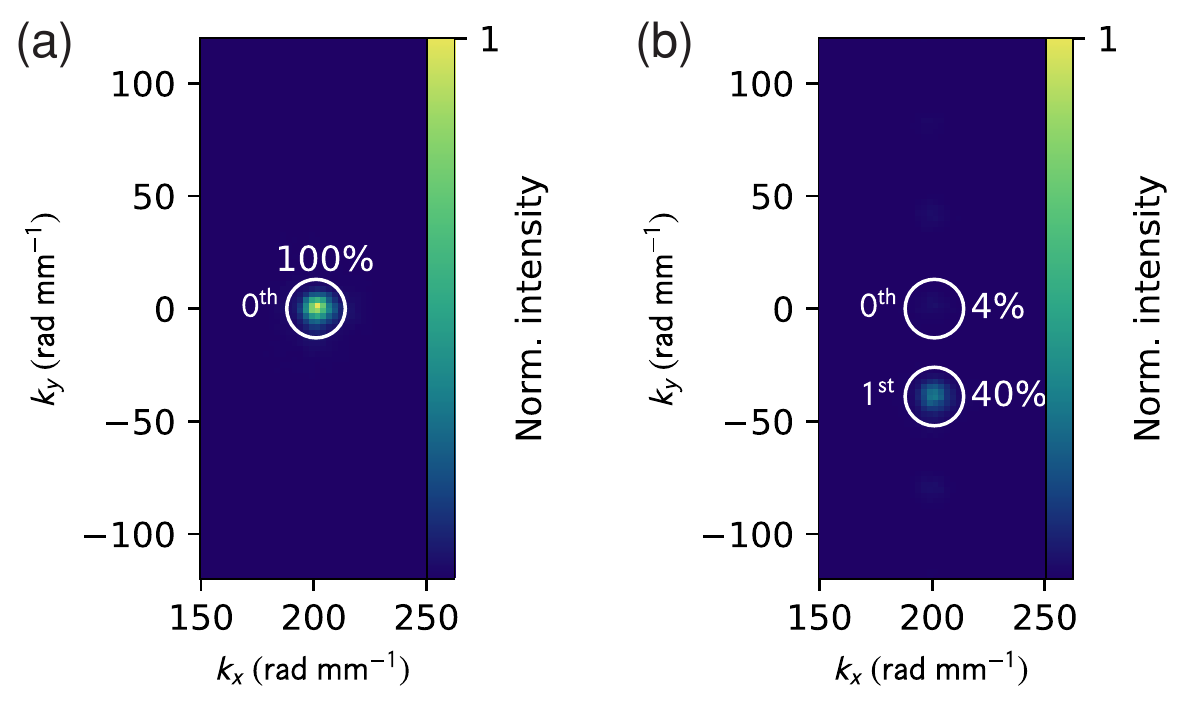}
	\caption{Wavevector-resolved light intensity emitted from a spin wave, (a) before and (b) after a ``blazed'' grating modulation is applied in order to transfer all excitations into the first diffraction order. \label{fig:rampe}}
\end{figure}
In Figs. 2(c) and 2(d) of the main text we presented that through changing the modulation pattern we may diffract spin-wave modes with desired amplitudes. As another example we demonstrate how to diffract one spin-wave mode into a single (distinct) mode. This can be simply achieved by selecting a linear ramp $\varphi_S(y)=\alpha y$, that transforms $\sw_{K_y}$ into $\sw_{K_y+\alpha}$. Such pattern however may require high laser intensities to significantly transform the mode. A good alternative is to wrap the phase shift ramp and obtain a ``blazed'' diffraction grating. The periodicity of this grating will determine the wavevector shift, and its amplitude is inherently $2\pi$. Fig. \ref{fig:rampe} presents an example manipulation with such grating, where we achieve 40\% transfer efficiency and only 4\% of excitations remain in the zeroth order. This figure of merit can be greatly improved if we can minimize current intensity deviations from a desired pattern of roughly 10\% by preparing a more uniform illumination of the spatial light modulator and designing a system with higher imaging resolution (see Section S.II for more details).
\paragraph{Towards more universality}
At this point we are able to perform phase operations in the real space (imprinting $\varphi(\mathbf{r})$), which correspond to a quite broad range of operations in terms of wavevector-space modes. To be able to perform arbitrary mode transformation we would need however more control. In particular, it has been recently shown in the context of frequency-bin modes  \cite{Lukens2017,Lu2018}, that an arbitrary mode transformation can be realized if we were able to imprint a phase independently onto distinct wavevector-space modes in between two real space phase imprints that we already perform. We envisage that this can be efficiently realized using a pair of atomic ensembles with a lens in between that transforms far-field coordinates of one ensemble onto near-field coordinates of the other ensemble. The protocol would be then to: (1) modulate spin wave in the first ensemble with $\varphi_1(\mathbf{r})$, (2) transfer them optically to the other ensemble situated in the far field, (3) effectively imprint the phase onto different wavevector-space modes $\varphi_2(\mathbf{K})$, (4) transfer the excitations into the first ensemble again and (4) perform the final real-space phase imprint  $\varphi_3(\mathbf{r})$. With storage (write-in and read-out) efficiencies approaching 90\% \cite{Cho2016} the protocol could be efficient as a whole.
\paragraph{Wavevector-multiplexed photon source}

To give estimates of performance of the multi-photon generation protocol utilizing the wavevector-multiplexed quantum memory with switching through ac Stark modulation let us first consider a scenario where the goal is to generate an $l$-photon state. We use a set of photon-pair sources based on two-mode squeezing (in the case of quantum memory they correspond to wavevector-space modes), which we will denote $w$ and $r$ for write-out and read-out, that remain in low photon generation rate regime $p=\mathrm{tanh}(\xi) \ll 1$, where by $p$ we denote a probability to generate a photon pair. The single photon in the read-out mode is heralded by detecting a photon in the write-out mode, and its quality is certified by a low value of second-order correlation function:
\begin{equation}
g^{(2)}_{r,r|w}=\frac{2p(2+p)}{(1+p)^2}\ll1
\end{equation}
We will denote transmission in the two modes by $\eta_w$ and $\eta_r$. Without a quantum memory we need $l$ of such sources. Each of them can herald a single photon with probability $p\eta_w$. Thus, the probability to herald $l$ photons will be $(p\eta_w)^l$.
With a quantum memory that harnesses $M$ modes we adopt a different strategy. Each of these modes now generates a photon-spin-wave pair with probability $p$. To be able to herald $l$ photons, and then redirect them into a desired spatial mode upon read-out, we require that the setup generates at least $l$ photons. For the switch we envisage to use our ac Stark modulation scheme. The efficiency of the transformation is taken into account in the effective read-out efficiency $\eta_r$. With the current scheme we can direct each of the $l$ photons in a desired mode, but only with a probability of roughly $1/l$ (since the grating is designed to operate as an inverse $l$-way splitter), so the success probability for all photons becomes $l^{-l}$. With the more universal scheme presented above such deficiency would be lifted.

\begin{figure}
	\begin{flushleft}
		(a)
	\end{flushleft}
	\centering
	\includegraphics[width=1\columnwidth]{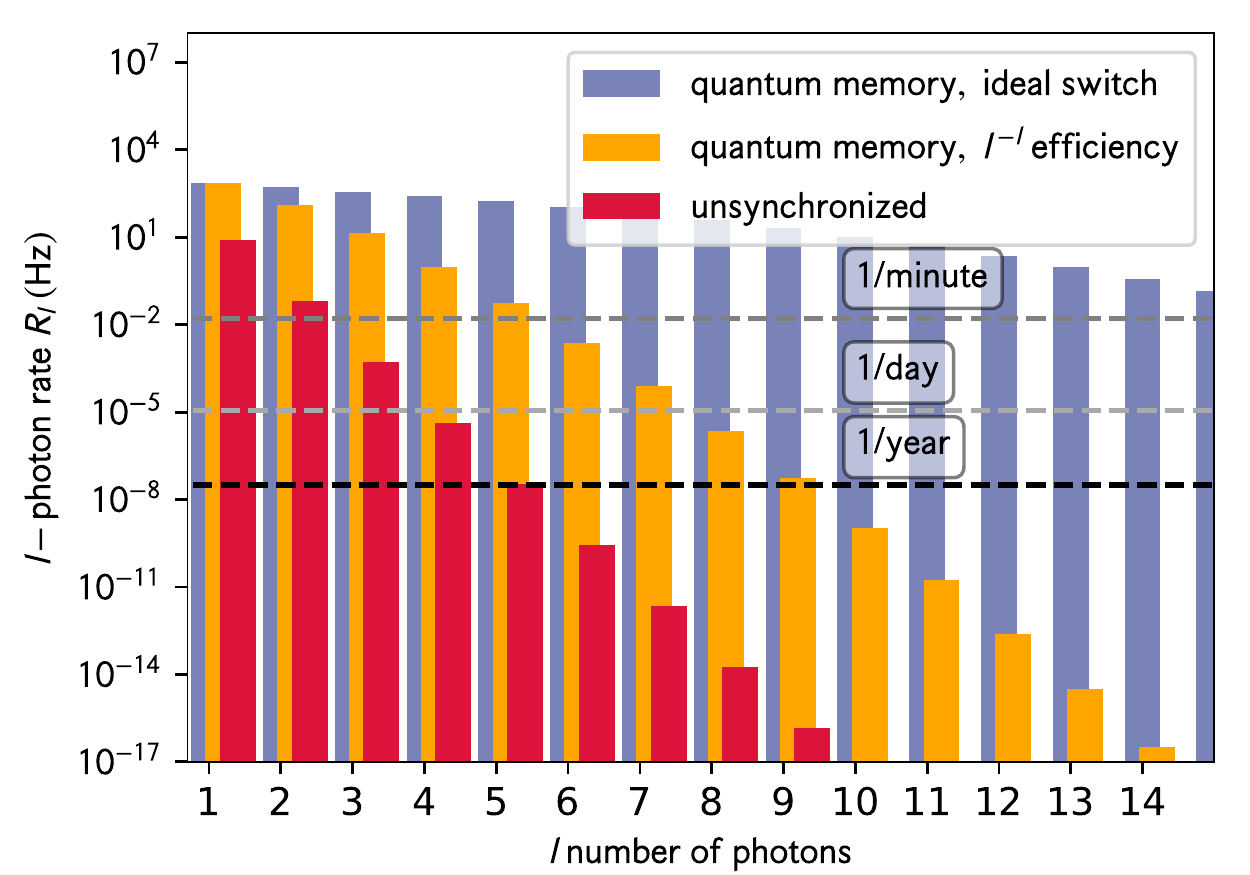}
	\begin{flushleft}
		(b)
	\end{flushleft}
	\includegraphics[width=1\columnwidth]{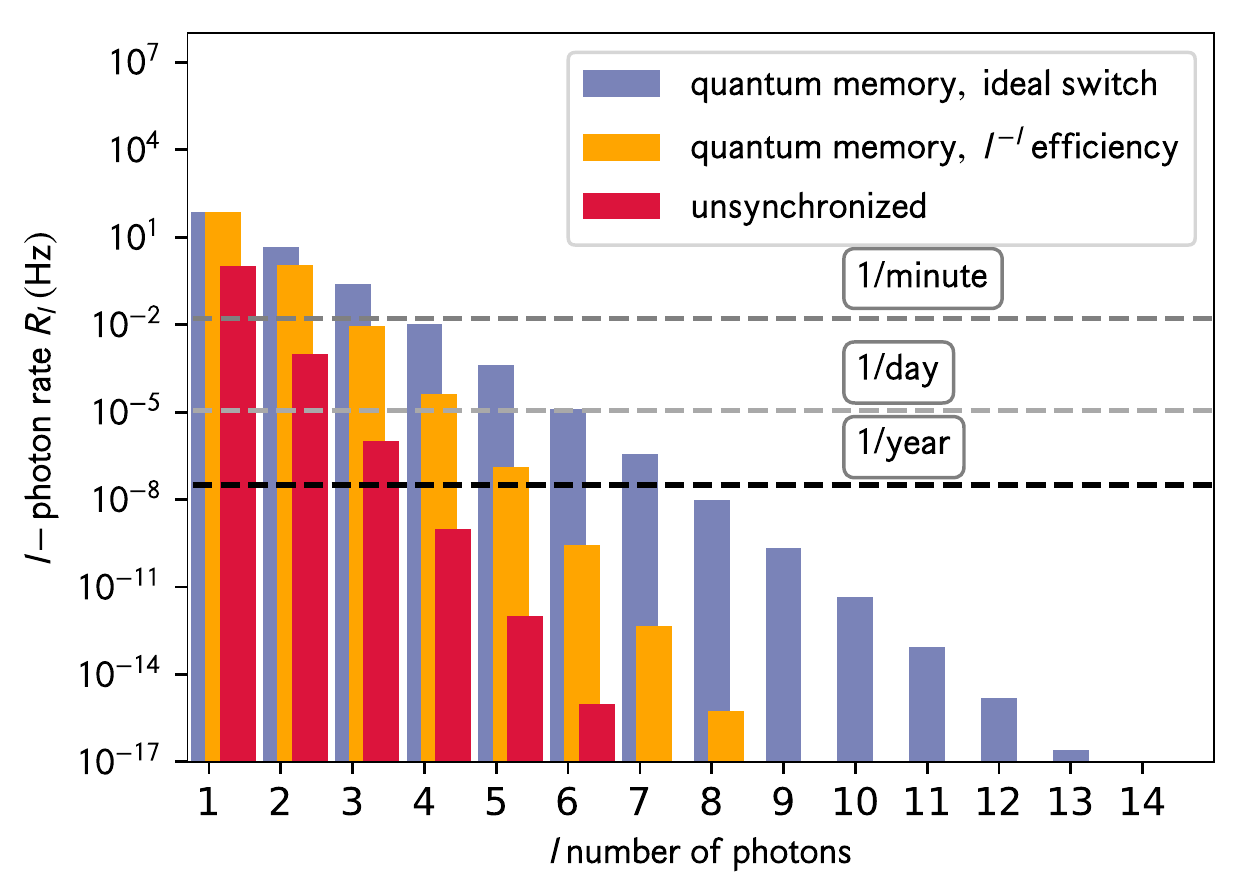}
	\caption{Estimates of the $l$-photon generation rates in two scenarios with a set of SPDC sources [$p=10^{-2}$, detection efficiency 90\% for (a) and 50\% for (b)] and a multimode quantum memory source [$p=10^{-2}$, number of modes $M=4000$, write photon detection (heralding) efficiency 20\%, read photon detection efficiency 90\% for (a) and 50\% for (b), effective read-out efficiency 80\% for (a) and 24\% for (b)] and repetition rate of 1 kHz. \label{fig:rata}}
\end{figure}

The probability to herald exactly $l$ photons is:
\begin{equation}
p_l = {M \choose l} (p\eta_w)^l (1-p\eta_w)^{M-l}
\end{equation}
which is a binomial distribution. The probability to herald at least $l$ photons is given by its cumulative distribution function:
\begin{equation}
p_{\geq l} = \sum_{j=l}^M {M \choose j} (p\eta_w)^j (1-p\eta_w)^{M-j} = I_{p\eta_w}(l,M-l+1),
\end{equation}
where $I_x(a,b)$ is a regularized incomplete Beta function.
Finally, in both cases we have losses in the idler arm which multiply our rates by $(\eta_r)^l$, and thus they equal (\emph{us} - unsynchronized, \emph{qm} - quantum memory):
\begin{gather}
P^{us}(l) = (p\eta_w \eta_r)^l \\
P^{qm}_M(l)=I_{p\eta_w}(l,M-l+1)(\eta_r)^l
\end{gather}
Via simple numerical studies one finds that in general it is optimal to guarantee a number of modes $M \gtrsim l(1+3/\sqrt{l}) / p \eta_w$ to obtain $I_{p\eta_w}(l,M-l+1)\approx1$, where $l/p \eta_w$ is a intuitive factor that simply describes that we need \textit{per-mode} probabilities to add up to one. The correction factor $3/\sqrt{l}$ reflects the width on which the considered Beta function rises (or more precisely 3 standard deviations of a distribution obtained by differentiation) and ensures that the probability of $l$-photon emission is already close to 1 rather than $\sim0.5$ as in the case without this factor. We require 3 standard deviations which ensures $I_{p\eta_w}(l,M-l+1)>0.98$.

Finally, we make a direct comparison with a spontaneous parametric down-conversion (SPDC) source and give optimistic, yet still realistic, estimates of expected photon rates. We optimistically assume detection efficiency of single-mode detectors equal 90\%, of camera detector equal 20\% and quantum memory read-out and transformation efficiency of 80\% with $M=4000$ modes and $p=10^{-2}$.  The low detection efficiency of the camera detector is compensated by the large number of modes. We take the repetitions rate of 1 kHz that corresponds to the best camera frame rate we can currently achieve. With these parameters we plot the expected $l$-photon rate $R_l$ in all scenarios in Fig. \ref{fig:rata}. We also compare these optimistic results with parameters closer to a current experiment (30\% readout and 80\% modulation efficiencies). Even in the case where the effective read-out efficiency is reduced by a factor of $l$, we obtain a significant enhancement (see Fig. \sref{fig:rata}{b}) over an unsynchronized source in all memory-enhanced cases.

An important experimental challenge remaining to realize the protocol is development of a real-time feedback sequence to prepare an appropriate ac Stark modulation pattern based on measured write-out photons. Current technology approaches this capability with new SLMs and digital micromirror devices (DMDs) approaching $\mu$s response times \cite{Andersen2014} and new fast and high-quantum-efficiency camera sensors \cite{Ma2017}. Ac Stark grating could also be generated with beams steered and interfered with the help of acousto-optic modulators.

\paragraph{Wavevector-multiplexed quantum repeater protocol}
Here we elaborate on the protocol for quantum entanglement distribution introduced in the main text. The new idea presented here relies on the possibility to generate DLCZ-type entanglement similarly as in the enhanced photon generation protocol described above. In particular, with a large number of modes $M$ we can increase the probability of pair generation $p$ until a desired number of photons reaches another node of a quantum network. At some point however, when distance is large and so are the losses, $p$ would have to become too large and we introduce a \textit{quantum repeater} in between the nodes. See Ref. \cite{Collins2007} for a related proposed multiplexed scheme.

\begin{SCfigure*}
	\centering
	\includegraphics[width=0.7\textwidth]{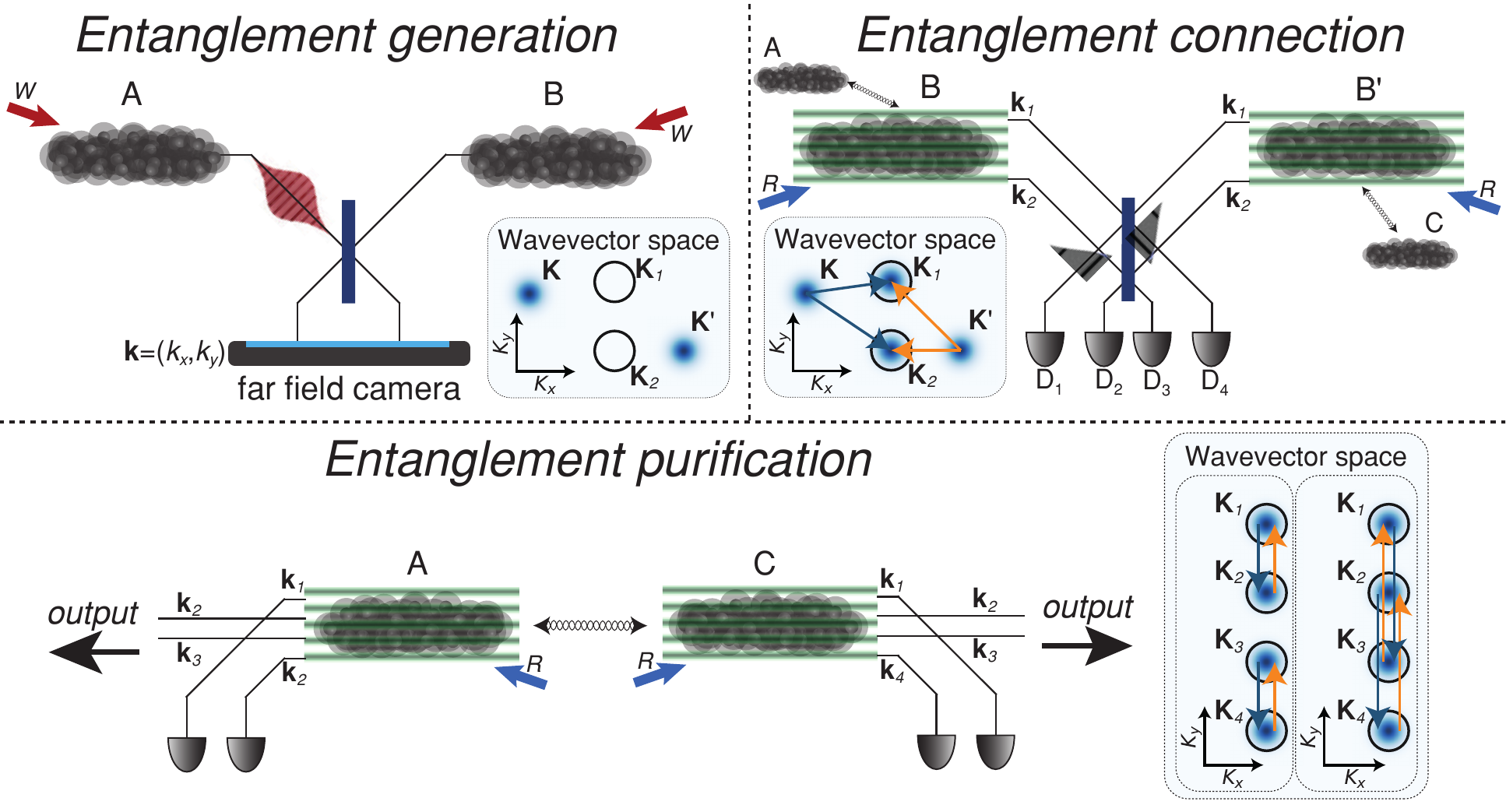}
	\caption{The three steps of the proposed quantum repeater protocol. \textit{Entanglement generation}: Two multimode quantum memories conneted by quantum channel are used to generate two entangled pairs of wavevector space encoded qubits. \textit{Entanglement connection}: At each repeater node spin wave modes are aligned and beamsplitter transformation generated by ac Stark modulation is applied. The read-out photons are combined on physical beamsplitter and detected by APDs. Proper coincidence pattern heralds successful connection. \textit{Entanglement purification}: Two previously connected pairs can be used to obtain one high fidelity entangled pair in purification process based on ac Stark beamsplitter transformation and coincidence measurement. \label{fig:sciema}}
\end{SCfigure*}

Our entanglement distribution protocol extends the proposals from \cite{Jiang2007} by employing two multimode quantum memories at each network node. Following these works the protocol begins with \textit{Entanglement generation} (ENG) between pairs of nodes (see Fig. \ref{fig:sciema}).  In this process two parallel entangled qubit pairs are generated, but in our case the pairs are sharing just two atomic ensembles and we use wavevector-space to encode memory qubits.
To explain this process let us consider two sites A and B connected with spatially multimode channel of length $L_{0}$. Both sites are excited simultaneously with write laser (W) to generate multi-mode multi-photon states. The photons are then combined on beamsplitter and detected by camera placed in far-field coordinatees of both ensembles. The multi-photon generation probability is set to ensure that overall two-photon detection probability $p_{\geq 2}$ with $\eta_w=\exp(-L_0/L_{att})\eta_{cam}$ (including detection efficiency $\eta_{cam}$ and channel loses with attenuation length $L_{att}$) is close to unity. Note that as in multiphoton-generation scheme thanks to multimode character of the used memories the probability $p_{\geq 2}$  can reach almost unity even for low \textit{per-mode} probability $p$ preventing unwanted single-mode multi-photon events. At each protocol iteration we register multi-photon counts on the camera. In the most simple scenario we then choose just two counts corresponding to detection of photons with wavevectors $\mathbf{k}$ and $\mathbf{k}'$. These events herald creation of spin waves with wavevectors equal $\mathbf{K}$  and $\mathbf{K}'$ distributed over both memories. After tracing out the remaining counts we obtain the two entangled qubit pairs encoded in the wavevector-space:
\begin{multline}
|\psi\rangle_{AB}^{ENG}=\frac{1}{2}\left( e^{i\phi}(|\mathbf{K}\rangle|\mathbf{K}'\rangle+|\mathbf{K}'\rangle|\mathbf{K}\rangle)\right.\\\left.+|\mathbf{K},\mathbf{K}'\rangle|vac\rangle+e^{2i\phi}|vac\rangle|\mathbf{K},\mathbf{K}'\rangle\right),
\end{multline} 
where $|\mathbf{K}\rangle\equiv \swd_{\mathbf{K}}|vac\rangle$. The tensor product corresponds to the two different atomic ensembles.

The next level of the protocol is \textit{Entanglement connection} (ENC). At this level we transform a pair of two entangled states $|\psi\rangle_{AB}^{ENG} \otimes |\psi\rangle_{B'C}^{ENG}$ (between memories $\mathrm{A}$ and $\mathrm{B}$, and $\mathrm{B}'$ and $\mathrm{C}$ respectively) to an entangled state between memory A and C:
\begin{equation}
|\psi\rangle_{AC}^{ENC} =\frac{1}{\sqrt{2}}\left(|\mathbf{K}_1\rangle|\mathbf{K}_2\rangle+|\mathbf{K}_2\rangle|\mathbf{K}_1\rangle\right).
\end{equation}
At the beginning of the ENC procedure we use the ac Stark modulation to transform the previously prepared entangled spin-wave pairs to occupy predefined memory modes $\mathbf{K}_1$ and $\mathbf{K}_2$. The same operation also performs beamsplitter transformation between the two chosen modes. After that, at each node we readout the spin waves and interfere the read-out photons on physical beamsplitter (BS) and then perform a coincidence measurement using APDs (Fig. \ref{fig:sciema}). As in \cite{Jiang2007} only 4 terms in state  $|\psi\rangle_{AB}^{ENG} \otimes |\psi\rangle_{B'C}^{ENG}$ contribute to the following coincidences: $\mathrm{D}_1\&\mathrm{D}_2$, $\mathrm{D}_1\&\mathrm{D}_3$, $\mathrm{D}_4\&\mathrm{D}_2$, $\mathrm{D}_4\&\mathrm{D}_3$. Any of those events heralds creation of the desired entangled state $|\psi\rangle_{AC}^{ENC}$ between sites A and C up to local qubit operations.

To obtain high fidelity entangled pair $\{\mathbf{K}_2,\mathbf{K}_3\}$ from two pairs ($\{\mathbf{K}_1,\mathbf{K}_2\}$ and $\{\mathbf{K}_3,\mathbf{K}_4\}$) the \textit{Entanglement purification} procedure can be applied. The procedure is based on known experimental \cite{Pan2003} and theoretical \cite{Jiang2007} proposals which we translate to wavevector-encoded qubits. First we combine modes of each pair and then combine the pairs using beamsplitter transformations caused by specific ac Stark modulation. Then the resulting spin-waves are retrieved into photons and contribution from modes $\mathbf{K}_1$ and $\mathbf{K}_4$ is detected using APDs. Coincidence event of these detectors heralds succesfull purification, otherwise the whole procedure is repeated.

\paragraph{Superadditive classical communication}
The operations presented in the previous section suggest that a very similar framework can be used to implement a Hadamard operation on modes (not to be confused with a Hadamard quantum gate), that would allow encoding of binary phase-shift keying bits in the Hadamard code and detecting the collectively. Such procedure can provide photon information efficiency gains in classical communications \cite{Guha2011,Klimek2016}. Our memory could also enable such collective measurements with feedback to provide a new perspective for communication \cite{Banaszek2012,Becerra2013} and quantum metrology protocols that require adaptive or collective measurements \cite{Hou2018}. 

\paragraph{Connection with Rydberg schemes}
%An enhanced, multiplexed source of ultranarrowband single-photons
%\cite{Nunn2013,Parniak2017} could also be achieved using a single
%atomic ensemble - with generation, multiplexed storage and feedback
%processing stages all realized within the SW domain. Multiplexing with feedback-based ACS control could also enhance the heralded SW pair generation rate and signal-to-noise ratio for our demonstration of HOM interference in this Letter.

%Recent advances in frequency conversion demonstrate the ability of
%an atomic-ensemble based processing unit to couple to an existing
%telecommunication networks and other QMs \cite{Maring2017}, 
While hitherto experiments
with Rydberg excitations (spin waves) have been performed in spatially single-mode
regime \cite{Petrosyan2017}, extending the capabilities to the continuous-variable multidimensional space could serve as a full photon-coupled platform
for simple quantum information processing.
Quantum computation and simulation schemes within
such a system endowed with the spatial resolution could range from
direct nonlinear quantum gates \cite{Petrosyan2017}, through testing effective field theories \cite{Gullans2016,Jachymski2016}, to a plenitude of more elaborate scenarios involving formation of topological SW
states \cite{Maghrebi2015} to perform fault-tolerant computation
\cite{Kitaev2003}. We envisage that to bring the Rydberg schemes into the wavevector-multiplexed domain we would need a few times larger blockade radius than currently achieved $\sim20\ \mu\mathrm{m}$ to support many spin-wave mode. 
\subsection{Experimental setup}

\paragraph*{Wavevector-resolved detection}

\begin{figure}[b]
	\includegraphics[width=1\columnwidth]{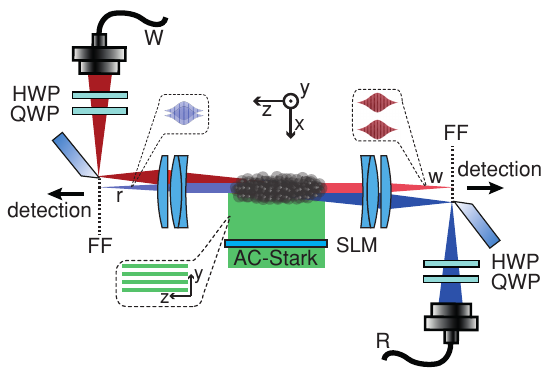}\caption{Schematic of the core of the experimental setup with marked write/read lasers and write-out/read-out photon modes. The write and read beams are counter-propagating and are separated from the generated photons in the far field (FF) of the atomic ensemble. The photons are collected from the MOT via a high field-of-view lenses. The ac Stark modulation beam shaped with a spatial light modulator (SLM) is applied from the side.\label{fig:ueksper}}
\end{figure}

\begin{figure}
	\includegraphics[width=1\columnwidth]{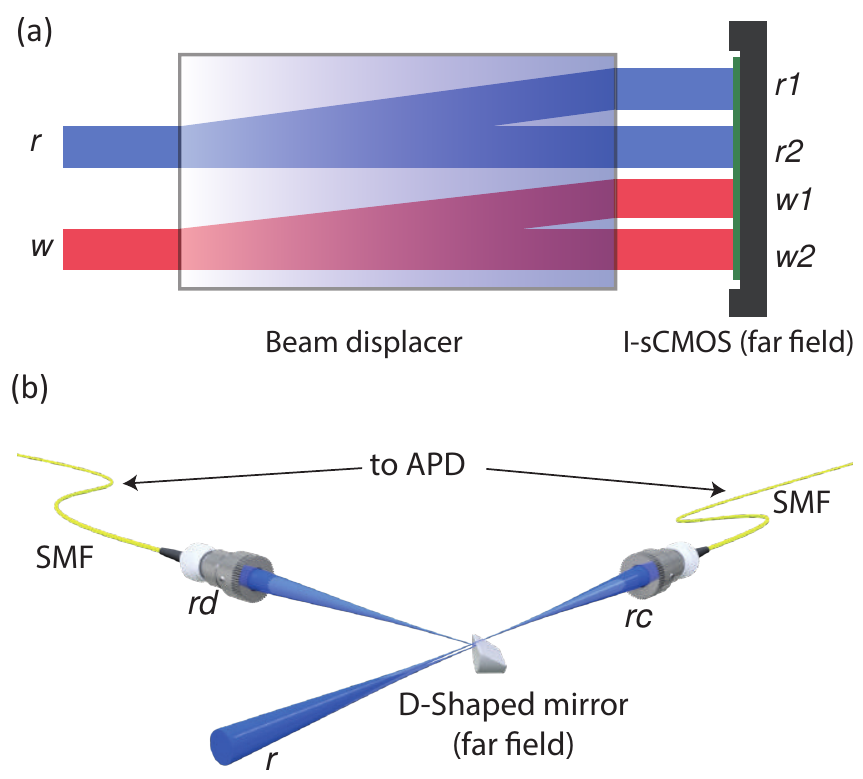}\caption{Two detection schemes used throughout the experiment. In
		panel (a) a two-dimensional camera sensor (photocathode of
		an image intensifier, I-sCMOS) situated in the far field with respect
		to the atomic ensemble detects single write-out ($w$) and read-out ($r$)
		photons in four distinct regions thanks to separation achieved with a calcite beam displacer. For cross-correlation measurements
		we add the photon counts from two regions corresponding to either
		write-out or read-out light. In panel (b) an example setup (one of
		the two used, for write-out and read-out photons) allowing detection of two
		far-field modes separated in $K_{x}$ and $K_{y}$. Single-mode fibers
		(SMF) collimators are aligned using XY translation stages. In terms of the wavevector-space fiber modes correspond to Gaussian shaped mode functions $u_\perp(\mathbf{K}_\perp)\propto e^{\frac{-|\mathbf{K}_\perp|^2}{4\sigma^2}}$. \label{fig:eksper}}
\end{figure}

A quantum memory based on an cold atomic ensemble prepared with
a magnetooptical trap (MOT), allowing wavevector-resolved detection
has been described in detail in \cite{Parniak2017}. See also Fig.~\ref{fig:ueksper} for a schematic of the core of experimental setup. An essential component of the detection setup is a spatially-resolved single photon detector comprising
an image intensifier based on a microchannel plate (Hamamatsu V7090D,
$\sim$20\% quantum efficiency) coupled with an fast and sensitive
scientific complementary metal-oxide semiconductor (sCMOS) sensor
(Andor Zyla 5.5 MP). Principles of operation at the single-photon
level and localization of single-photon flashes are detailed in Refs.
\cite{Chrapkiewicz2016,Dabrowski2017,Chrapkiewicz2014,Lipka2018}. Here, before the camera, write-out
and read-out photons are separated into two regions to allow HBT measurement
[see Fig. \sref{fig:eksper}{a}]. Using separate regions of the camera
effectively allows photon-number resolved detection \cite{Chrapkiewicz2014} without deleterious cross-talk effects \cite{Lipka2018}. Photons emitted from the cold atomic ensemble are imaged in
the far field onto the image intensifier using a complex multi-lens
setup (effective focal length of 50 mm) with angular resolution of
0.6 mrad \cite{Dabrowski2017}, corresponding to a wavevector of $4.7\ \mathrm{rad\ mm}^{-1}$ (see Fig. \ref{fig:ueksper}).
Optically-pumped filtering cells heated to $60^{\circ}\mathrm{C}$
containing Rubidium-87 and Krypton (1 Torr pressure) as buffer gas
are used to separate stray laser light from single Raman-scattered
photons. Typical measurement comprises $10^{7}$ camera frames. Note
that spatially-insensitive filtering is crucial for this experiment,
as typically used filtering cavities would allow us to use only a
single spatial mode, negating the prospects of scalability. For the
experiment involving highly populated classical spin-wave states we
used the same detector operated in the proportional-intensification
regime achieved by lowering the electron gain of the image intensifier
(see Ref. \cite{Dabrowski2014} for details on this operation regime).

For the quantum interference experiment we replace the intensified
sCMOS (I-sCMOS) camera with single-mode fibers coupled to single-photon
avalanche photodiodes (APD, Perkin Elmer, $\sim$50\% quantum efficiency).
Fiber detection modes [see Fig. \sref{fig:eksper}{b}] correspond to Gaussian beams with waist radius
of 0.15 mm centered in the atomic cloud. The APDs allow for faster
experimental repetition rate and provide higher quantum efficiency
than the image intensifier. Coherent spatial filtering with single-mode
fibers additionally mitigates the requirement of very narrow post-selection
of wavevectors (or positions, like in the atomic experiment of \emph{Lopes
	et al.} \cite{Lopes2015} that used a microchannel plate) in the HOM
experiment.

\paragraph*{Shaping of the ac Stark beam}

Precise shaping of the ac Stark beam is essential to obtain the desired
effect. One reason is the need to obtain a desired pattern, but very
importantly a constant intensity profile along the $z$ direction,
possibly free of distortions and inhomogeneities \cite{Leszczynski2018},
is required to read-out spin waves efficiently. Otherwise atoms in
different places along the $z$ direction will accumulate random phases,
resulting in poor phase matching at the read-out stage. For accurate
shaping of the beam we use a spatial light modulator (SLM, Holoeye
Pluto) coupled with a charge-coupled device (CCD) camera (Basler Scout
scA1400-17fm). The SLM is illuminated with an elliptically shaped
beam from a semiconductor taper amplifier (Toptica, BoosTA)
seeded with a light from an ECDL (Toptica DL 100) locked using an
offset-lock setup \cite{Lipka2016}. Temporal ac Stark pulse profile is controlled with an
acousto-optic modulator. Both the CCD camera and the atomic ensemble
are situated in the same image plane of the SLM ($\times1.7$ magnification).
Importantly, same lenses are used and the only difference between
the camera plane and the atomic cloud plane is a flip mirror instead
of a vacuum chamber window on the beam path. With this, we achieve
best possible representation of light intensity in the vacuum chamber,
distorted by a minimal number of optical elements. Any spatial noise of variance along the longitudinal $z$ direction $\mathrm{Var}_\mathbf{z}{(\varphi)}$ then translates to reduced spin wave retrieval efficiency scaling as $\mathrm{exp}(-\mathrm{Var}_\mathbf{z}{(\varphi)}/2)$ in a benchmark scenario of flat pattern illumination, where the statistics of intensity noise are found to be Gaussian. This noise tends to deviate significantly from a Gaussian shape when a grating is prepared, thus we rather choose to adopt a phenomenological model of exponential decay $\mathrm{exp}(-\gamma \chi)$, where $\chi$ is the modulation depth and $\gamma$ an associated decay constant.

To generate the desired light intensity profile, we first map the
camera pixels onto SLM pixels, taking into account possible rotation
and distortions. Next, we use an iterative algorithm with feedback
from the CCD camera to generate a desired pattern. Due to phase flicker,
the experimental sequence, both during the experiment and during the
calibration and optimization of the ac Stark beam profile, is synchronized
with the SLM refresh rate.

\begin{figure}
	\includegraphics[width=0.9\columnwidth]{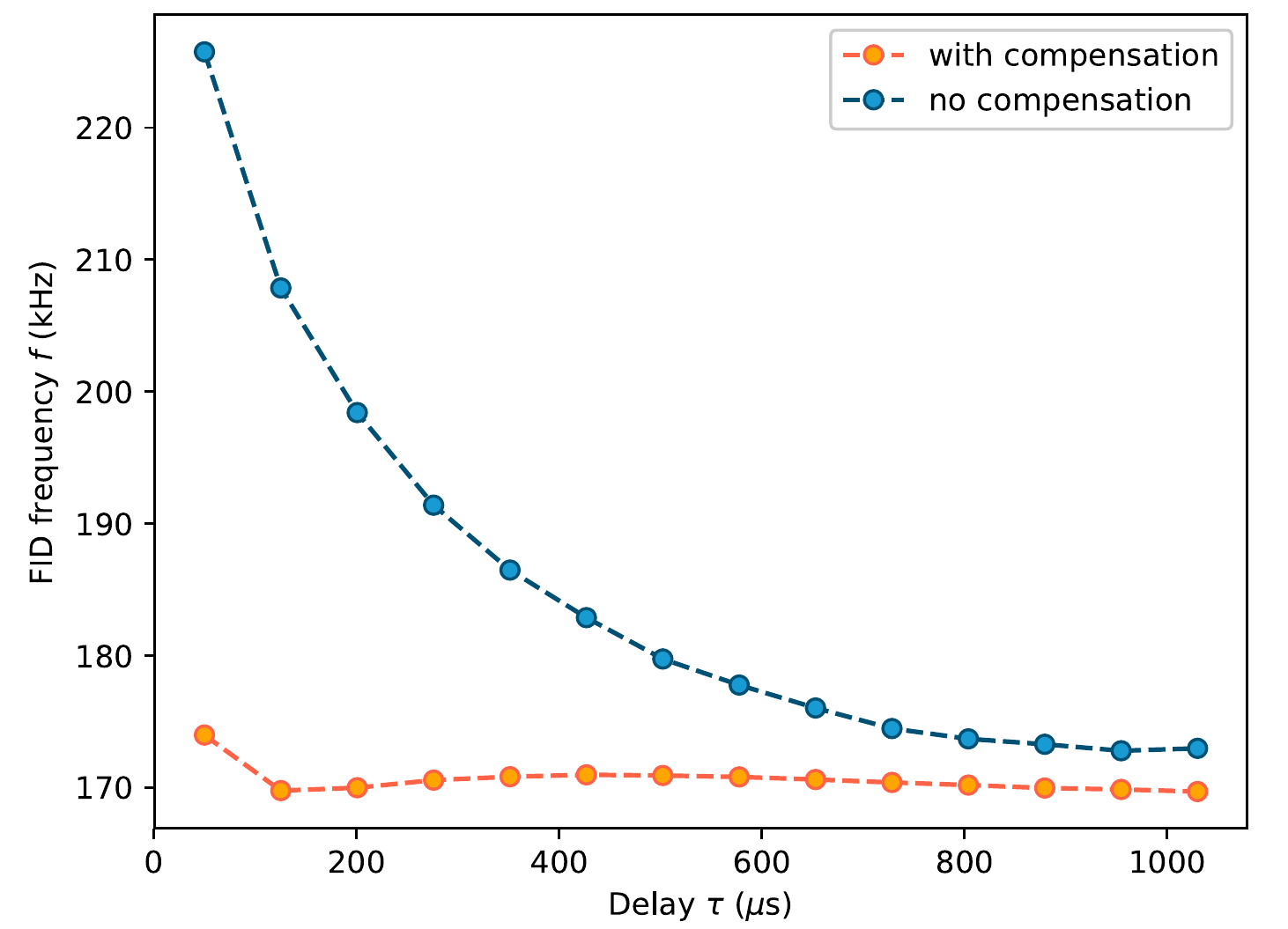}\caption{Compensation of magnetic fields induced by eddy currents created after switching off the MOT coils. For this measurement we apply a small transverse instead of longitudinal magnetic field to obtain a strong spin-precession signal by probing atoms with linearly polarized light (FID, free-induction decay). The atoms are first prepared in the $F=1,\ m_F=1$ state. We than apply 50 $\mu$s long probing sequences. Before each sequence atoms are again optically pumped. From each sequence we extract the central oscillation frequency (proportional to the magnetic field with a constant 1.4 kHz/mG) and plot it as a function of delay after the MOT coils are switched off. We observe that without a shorted coil the spurious fields change on a 0.5 ms timescale, while an additional coil guarantees their rapid stabilization. \label{fig:compens}}
\end{figure}

\begin{figure*}
	\includegraphics[width=1\textwidth]{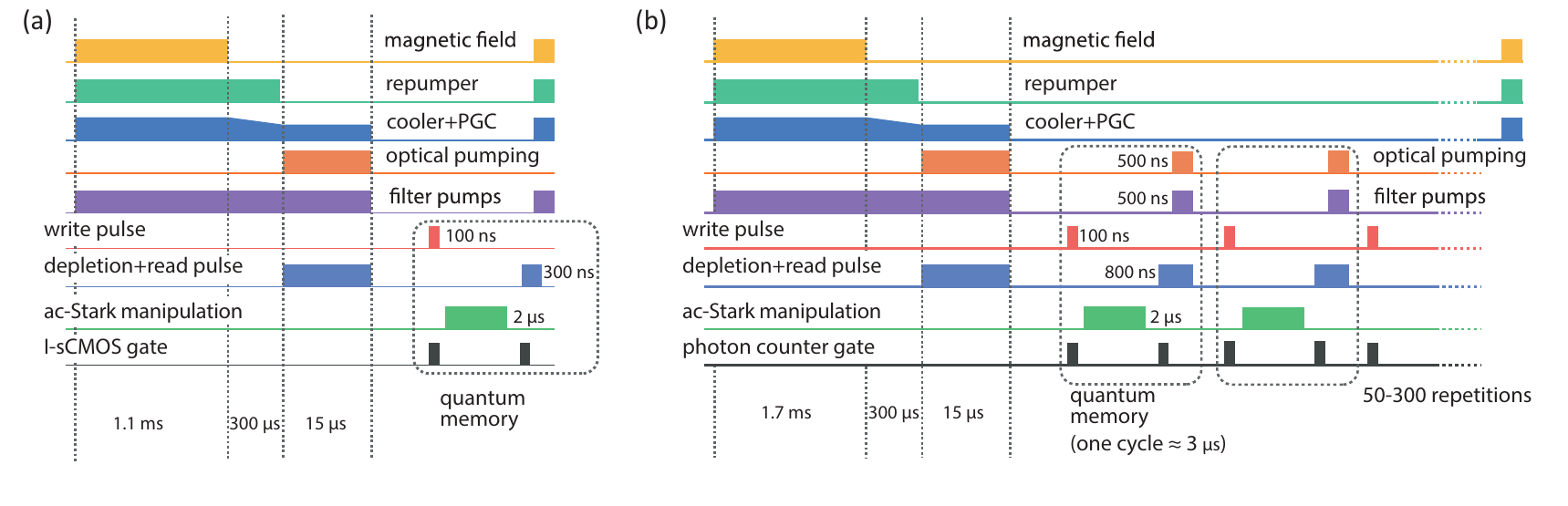}\caption{Timing sequence of the experiment. Panel (a) portrays
		the timing sequence used when wavevector-resolved detection using
		an I-sCMOS camera is performed. Due to camera frame rate limitation
		only one cycle of QM per one MOT loading is performed. In (b) we show
		the timing sequence for experiment using few-mode detection using
		APDs. One MOT cycle fits up to 300 QM cycles. In both cases the sequence
		is repeated at the rate of 420 Hz.\label{fig:tajming}}
\end{figure*}

\begin{SCfigure*}
	\includegraphics[width=0.65\textwidth]{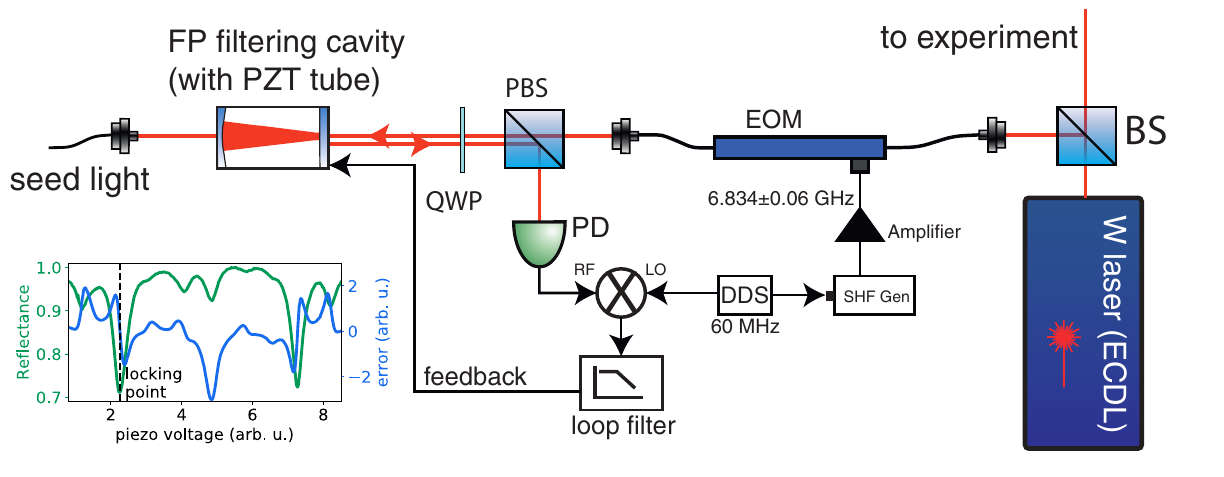}\caption{Generation of seed light. To generate phase-coherent light
		detuned from the write laser by exactly the Rubidium-87 hyperfine
		splitting, we send some of the write laser light into an electro-optic
		modulator (EOM) producing sidebands at $f_{\mathrm{SHF}}=\pm6.834$
		GHz. A Fabry-P\'erot (FP) scanning cavity is used to filter out all
		sidebands except the desired one. For this, we additionally modulate
		the $f_{\mathrm{SHF}}$ frequency at $f_{\mathrm{RF}}=60$ MHz. Light
		reflected from the cavity is registered using a photodiode (PD) and
		the signal is mixed with the $f_{\mathrm{RF}}=60$ MHz modulation,
		producing a locking signal for the cavity (inset).\label{fig:sidowanie}}
\end{SCfigure*}

\paragraph*{Sequence}

Fig. \ref{fig:tajming} presents the timing sequence of the experiment,
used both in the configurations based on detection with I-sCMOS (panel
a) or APDs (panel b). The sequence is repeated at the 420 Hz refresh
rate of the SLM in both cases. In a single cycle, the atoms are trapped
for 1.8 ms in the case of the I-sCMOS experiment (1.1 ms in the case
of the APD experiment). With a relatively high Rb vapor pressure ($\sim10^{-7}$ mbar) and cooling laser (70 mW total power) as well as a detuning of 15 MHz to the red from the $5^2S_{1/2}\ F=2\rightarrow 5^2P_{3/2}\ F=3$ transition the sequence allows us to maintain a stable atom number in the MOT. The number of atoms saturates after only approx. 10 s that include thousands of MOT cycles, yielding an optical depth of $30$ up to $200$ (strongly depending on how many times the quantum memory cycle is repeated), as measured at the $5^2S_{1/2}\ F=2\rightarrow 5^2P_{3/2}\ F=3$ closed transition of the D2 line. Trapping is followed by polarization gradient
cooling in optical molasses (PGC) with the cooling laser detuning increased to 31 MHz that allows us to reach a temperature
of $22\ \mu\mathrm{K}$. As currently we store spin waves for only $2\ \mu\mathrm{s}$, the temperature is not essential for obtained results. The trapping magnetic fields produced with 125 A current in a low-inductance MOT coil (designed similarly as in \cite{Zhang2012}, supporting gradients of 25 G/cm in the MOT) are switched off to allow
the eddy currents to decay. A switch based on MOSFET transistors turns off the current in less than 5 $\mu\mathrm{s}$. An additional shorted coil is placed above the vacuum chamber to compensate for the eddy currents in the optical table, which yields a stable magnetic field at 100 $\mu\mathrm{s}$ after the MOT coil is switched off. 

Additional large coils around the entire setup allow setting a constant bias field. To compensate the eddy currents we performed an additional experiment and set a small bias field in the $x$ direction, so that we obtained a strong spin-precession signal, as the atoms are $z$ polarized with an optical pump. By measuring the free-induction decay signal we observed that indeed our setup allows rapid decay of stray magnetic fields after MOT coils are turned off (see Fig. \ref{fig:compens}). During operation of the quantum memory, that is for the proper experiment, we set a 50 mG bias magnetic field along the $z$ direction, which allows better optical pumping and selection of a proper magnetic sublevel.

The atoms are finally prepared in
the $F=1$, $m_{F}=1$ state through optical pumping with $>70\%$ efficiency, using one laser (15 mW power) tuned to the $5^2S_{1/2}\ F=2\rightarrow 5^2P_{1/2}\ F=2$ transition and another circularly-polarized laser (10 mW power) tuned to the $5^2S_{1/2}\ F=1\rightarrow 5^2P_{3/2}\ F=1$ transition.

A single QM
cycle consists of a $100\ \mathrm{ns}$ long write pulse (varying power, typically $\sim2\ \mu\mathrm{W}$), a $2\ \mu\mathrm{s}$
long ac Stark spin-wave manipulation pulse, and 300 ns read laser pulse (300 $\mu\mathrm{W}$ power). The
write pulse is left-circularly polarized and red-detuned by
25 MHz from $5^{2}S_{1/2},\ F=1\rightarrow5^{2}P_{3/2},\ F=2$ transition). 
The counter-propagating read laser pulse is right-circularly polarized and resonant with $5^{2}S_{1/2},\ F=2\rightarrow5^{2}P_{1/2},\ F=2$
transition. See also Fig. \ref{fig:ueksper} for experimental geometry. All lasers are locked to either cooler or
repumper laser through a beat-note offset lock \cite{Lipka2016}.
In the I-sCMOS experiment the image intensifier gate is open during
writing and reading. The sCMOS camera captures photon flashes during
both gates, in separate spatial regions of the image intensifier.
In the APD experiment the memory cycle is followed by a short 500
ns clear pulse (consisting of read laser pulse, optical pumping and
additional pumping of filtering cells to maintain hyperfine polarization). The APD gate is kept open only during the initial 80 ns of the read pulse. In this way we optimize the ratio of signal to dark counts, that still comprise 20\% of the detected read-out photocounts.

A single QM cycle can be repeated up to 300 times per one MOT. Within this configuration and in the modulation-free experiment we typically detect approx. 40 coincidences per second in a single pair of modes, with $g^{(2)}_{wr}\approx20$ and write-out photon detection probability of $10^{-2}$ in the write-out mode. This translates to 1.5 quadruple coincidences per minute. When the three-way splitter modulation is applied, the quadruple coincidence is inherently reduced to less than 0.5 per minute. With lower repetition rate we obtain a larger MOT (with OD up to 200) and better efficiency, yet the rate of quadruple coincidences is slightly reduced to roughly 0.1 per minute.

A faster photon pair generation rate can be achieved using an FPGA-based
feedback, in which we only applied ac Stark manipulation and read laser
pulse if a write photon (or a pair of write photons) is detected.
While such configuration nearly triples the generation rate, it provides
less data as not all correlation functions can be tracked this way.
Nevertheless, this is a recommended operation scheme for future experiments
with spin-wave pairs.

An alternative method for improving the generation rate would be to utilize an atomic ensemble with higher optical depth to increase the read-out efficiency. Photodetectors with higher quantum efficiency and lower dark count rates would also significantly improve the observed experimental parameters.
\paragraph*{Generation of coherent spin-wave states}

To generate a highly-populated coherent spin-wave state we seed the
Stokes scattering process. The process is governed by a squeezing
Hamiltonian so both the seed light and the spin waves are amplified;
however, for strong and coherent seed light, the generated spin-wave
state is close to a coherent state. The seed light, detuned by the
Rubidium-87 hyperfine splitting from the write laser light, needs to
be phase-coherent with the write laser for the process to be efficient.
We use an electro-optic modulator (EOM) fed with an super-high-frequency (SHF) signal with
central frequency $f_{\mathrm{SHF}}=6.834$ GHz to generate sidebands
(see Fig. \ref{fig:sidowanie} for experimental schematic).
The SHF signal is additionally modulated using a 60 MHz sine wave
from a direct digital synthesizer (DDS). Modulated light consisting
of harmonic frequency components separated by $f_{\mathrm{SHF}}$
is sent to a FP-cavity and its reflected portion is directed onto
a fast photodiode. The photodiode registers beat-notes (RF) at 60
MHz, which are then mixed with the 60 MHz local oscillator (LO) signal.
After the loop filter, we obtain a locking signal, which thanks to
a proper choice of relative phases between LO and RF allows locking
only at the desired sideband (the locking signal slopes for positive
and negative shifts differs in sign). For seeding purposes we are
interested only in the term which is shifted by $-f_{\mathrm{SHF}}$
from the original laser frequency. The cavity reflects the fundamental
unmodulated light and other sidebands, resulting in 26 dB net attenuation
of all unwanted components. Importantly, this novel setup allows generation
of very pure seed light with only one modulator and an uncomplicated
cavity system (cf. \cite{Hosseini2009,Bernien2017})

\subsection{Phase modulation with the ac Stark beam}
\begin{figure}
	\includegraphics[width=1\columnwidth]{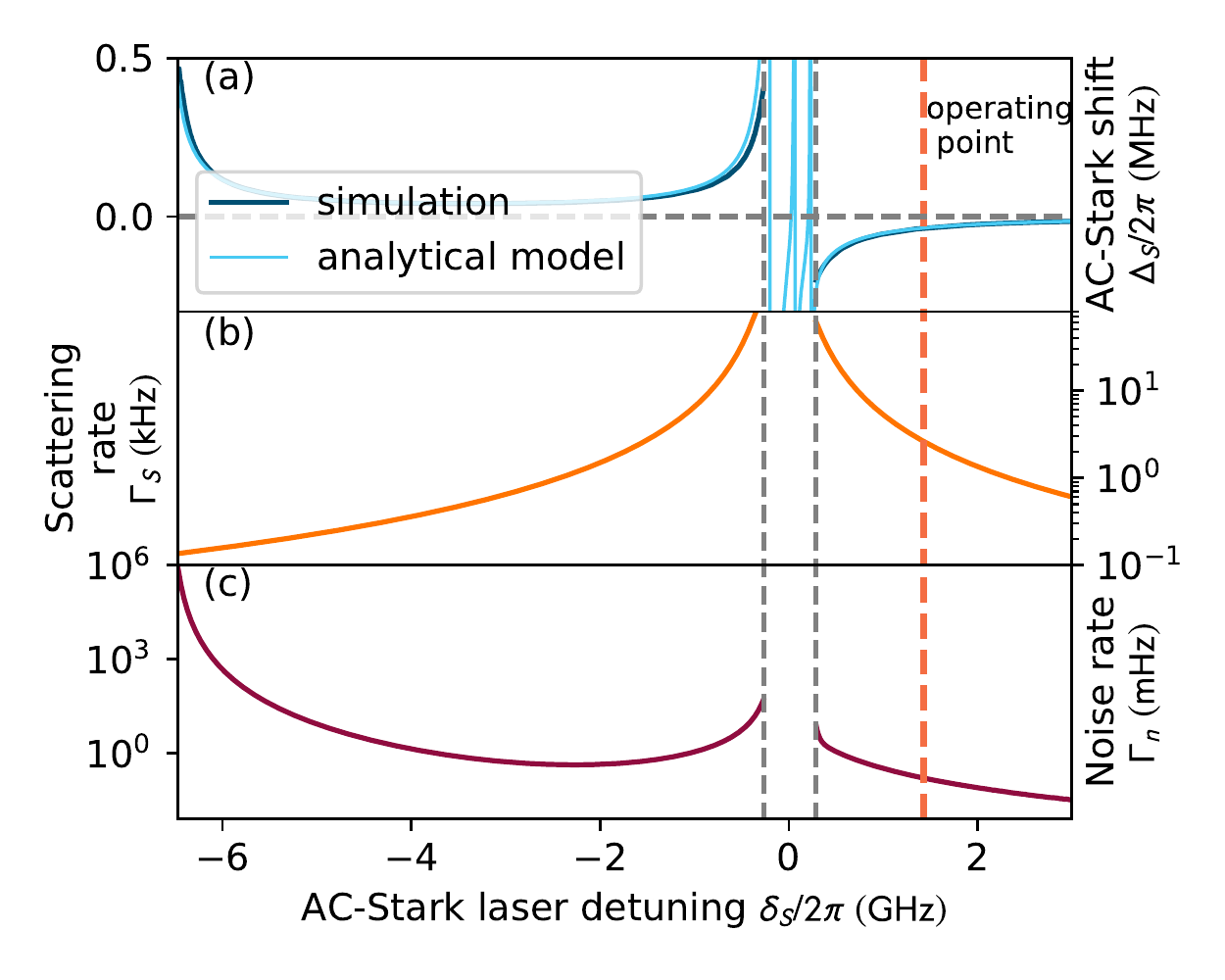} 
	
	\caption{Properties of the ac Stark modulation. Panel (a) depicts
		the differential ac Stark shift for the Rabi frequency of the ac Stark
		beam equal with an intensity of $35\mathrm{\ mW/cm^2}$, as a function of ac Stark laser detuning $\delta_{S}$
		from the $5^{2}S_{1/2}F=2\rightarrow5^{2}P_{3/2}$ transition
		centroid. Both the simulation and the simplified analytical model are presented. A small deviation between the two can be noticed only close to resonance, as decoherence is not taken into account in the simplified model. Panels (b) and (c) portray incoherent scattering
		rate (decoherence rate) of spin-waves and generation rate of spurious
		excitations, respectively. The values are not plotted very close to
		the absorption resonance due to very high incoherent scattering rendering
		this region useless for the purpose of ac Stark modulation.\label{fig:odstrojenia}}
\end{figure}

To theoretically evaluate the performance of the ac Stark modulation
at the single-excitation level, it is crucial to consider both the
level of decoherence caused by manipulation and the spurious noise
produced. The manipulation should also influence the coherence in
a proper way, i.e. the states $|g\rangle$ ($F=1$, $m_{F}=1$) and
$|h\rangle$ ($F=2$, $m_{F}=-1$) should be eigenstates of the effective
ac Stark splitting Hamiltonian \cite{Sparkes2010,Leszczynski2017}.
Otherwise, the spin wave is transferred to a different combination
of magnetic sublevels, which may lead to beat-notes as well as decoherence
due to magnetic field inhomogeneities. We found the optimal setting
is to red-detune the ac Stark laser by $\delta_{S}=0.5$\textendash $3.0$
GHz from the ``empty'' state $|h\rangle$ (we calculate the detuning
from $5^{2}S_{1/2}F=2\rightarrow5^{2}P_{3/2}$ transition centroid,
lying 193.7 MHz below the $F=2\rightarrow F=3$ transition),
so the energy shift of the $|h\rangle$ state is much larger than
for the $|g\rangle$ state. While this causes some scattering from
the $|h\rangle$ state, we avoid exciting atoms from the $|g\rangle$,
which could generate spurious spin-wave excitations that would later
be retrieved as noise. Furthermore, by making the ac Stark light $z$-polarized,
we ensure that $|h\rangle$ is an eigenstate of the effective ac Stark
shift Hamiltonian. Note that this setting is very different from the
proposal of \emph{Sparkes et al.} \cite{Sparkes2010} who considered
much larger detunings. In their setting, the ac Stark light is coupled
to $|g\rangle$ and $|h\rangle$ with nearly the same strength and
the differential phase shift only appears when circular polarizations
are used and in only several specific spin-wave magnetic configurations.
Such an operation requires multi-watt power levels to obtain reasonable
differential ac Stark shifts. Furthermore, while the scattering rate
would be indeed small, the noise generation rate has not been considered
and could become a significant problem.

To evaluate the above predictions we model the full behavior of the
multi-level atom described by a density matrix $\rho$ subject to
an off-resonant ac Stark field. We consider a full Hamiltonian $\boldsymbol{H}$
including all ground-state and excited-state sublevels and the following
master equation:

\begin{equation}
\frac{{\mathrm{{d}}\rho}}{\mathrm{{d}}t}=\frac{{1}}{i\hbar}[\rho,\boldsymbol{H}]-\frac{{1}}{2}\{\boldsymbol{\Gamma},\rho\}-\mathrm{{Tr}}(\rho\boldsymbol{F})+\boldsymbol{\Lambda}_{\rho},
\end{equation}

where $\boldsymbol{\Gamma}$ is the relaxation matrix corresponding to the ground-state manifold, $\boldsymbol{\Lambda}_{\rho}$
is the repopulation matrix corresponding only to the ground-state manifold as well, and $\boldsymbol{F}$ is the spontaneous
emission operator \cite{Auzinsh2010,Parniak2015}. First we prepare
a spin-wave state as $\left(|g\rangle+\epsilon|h\rangle\right)/\sqrt{{1+\varepsilon^{2}}}$
(with $\varepsilon\ll1$) and track the behavior of the atomic state
in the subspace spanned by $|h\rangle$ and $|g\rangle$. The relative
phase is calculated as $\varphi=\mathrm{Arg}(\rho_{hg})$ and the
ac Stark shift as $\Delta_{S}=\left.\frac{\mathrm{d}\varphi}{\mathrm{d}t}\right|_{t=0}$.
The scattering rate $\Gamma_{S}=-\varepsilon^{-2}\left.\frac{\mathrm{d}\rho_{hh}}{\mathrm{d}t}\right|_{t=0}$
quantifies the decay of the spin-wave population. For the noise rate $\Gamma_{n}$
we take an atom prepared in a pure $|g\rangle$ state and we again
calculate the rate as $\Gamma_{n}=\left.\frac{\mathrm{d}\rho_{hh}(\varepsilon=0)}{\mathrm{d}t}\right|_{t=0}$.
The results are plotted in Fig. \ref{fig:odstrojenia}
as a function of ac Stark laser detuning $\delta_{S}$. We find
the differential splitting $\Delta_{S}/2\pi=-0.036\ \mathrm{MHz}$ for the operation point at $\delta_{S}/2\pi=1.43$ GHz and with approx.
40 mW of average power corresponding to an intensity of $35\ \mathrm{mW/cm^2}$. Our experimental setup facilitates a peak power of 100 mW.

We can elucidate on these results by introducing a simple analytical model. In the model we include all transitions via which the ac Stark shift is induced in the far-detuned regime. The shifts of the two ground-state levels are:
\begin{multline}
\Delta_S ^{(|g\rangle)} = \frac{|\mathcal{E} d|^2}{4\hbar} \biggl( \frac{5}{24}\frac{1}{\delta_{S}+\frac{5}{4} A_{0,1/2} - \frac{11}{4} A_{1,3/2}} + \\
+ \frac{1}{8}\frac{1}{\delta_{S}+\frac{5}{4} A_{0,1/2} - \frac{3}{4} A_{1,3/2}} \biggr)
\end{multline}
\begin{multline}
\Delta_S ^{(|h\rangle)} = \frac{|\mathcal{E} d|^2}{4\hbar}\biggl(\frac{1}{40}\frac{1}{\delta_{S}-\frac{3}{4} A_{0,1/2} - \frac{1}{4} A_{1,3/2}} + \\
+ \frac{1}{24}\frac{1}{\delta_{S}-\frac{3}{4} A_{0,1/2} - \frac{3}{4} A_{1,3/2}}\\
+ \frac{4}{15}\frac{1}{\delta_{S}-\frac{3}{4} A_{0,1/2} + \frac{9}{4} A_{1,3/2}} \biggr)
\end{multline}
with $\mathcal{E}$ being the field of the ac Stark beam, $d=3.58 \times 10^{-29}\ \mathrm{Cm}$ being the transition dipole matrix element for the D2 line, and $A_{0,1/2}/2\pi=3.42\ \mathrm{GHz}$, $A_{1,3/2}/2\pi=85\ \mathrm{MHz}$ being the hyperfine coupling dipole constants. The differential ac Stark shift is calculated as:
\begin{equation}
\Delta_S = \Delta_S ^{(|h\rangle)} - \Delta_S ^{(|g\rangle)}
\end{equation}
In our scenario, the Rabi frequencies corresponding to subsequent transitions are: 13.0 MHz ($F=1\rightarrow F=1$), 10.1 MHz ($F=1\rightarrow F=2$), 4.5 MHz ($F=2\rightarrow F=1$), 5.8 MHz ($F=2\rightarrow F=2$) and 14.7 MHz ($F=2\rightarrow F=3$). Due to detuning, only the latter three transitions contribute significantly.
This setting provides a total phase shift of $\phi_{S}=\Delta_{S}T$
of the order of 0.45 rad with the manipulation time $T=2\ \mu\mathrm{s}$. We find the scattering
rate $\Gamma_{s}=390$ Hz, which results in destruction of less than
0.1\% of the spin waves due to incoherent excitation. Finally, we find a very
little noise generation rate per atom $\Gamma_{n}=1$ mHz (cf. with
significantly higher noise rate when ac Stark laser is tuned closer
to the $|g\rangle$ state at $\delta_{S}\approx-6.8$ GHz in Fig. \ref{fig:odstrojenia}). If we assume that photons from
these spurious excitations are scattered randomly during read-out
to all far-field modes the number of which we estimate as $\sigma_{z}\sigma_{\perp}^{2}/\lambda^{3}\approx7\times10^{8}$
with $\sigma_{z}=4\ \mathrm{mm}$, $\sigma_{\perp}=0.3\ \mathrm{mm}$
(longitudinal and transverse sizes of the ensemble, respectievly), $\lambda=795\ \mathrm{nm}$
(wavelength of the read laser) and $N=10^{8}$ (number of atoms),
we estimate the probability of emitting a noise photon per mode of
less than $3\times10^{-10}$, which is completely negligible compared
with e.g. noise introduced by imperfect optical pumping.

\subsection{Phase pattern design and spin-wave splitter performance}

\begin{figure}
	\includegraphics[width=0.9\columnwidth]{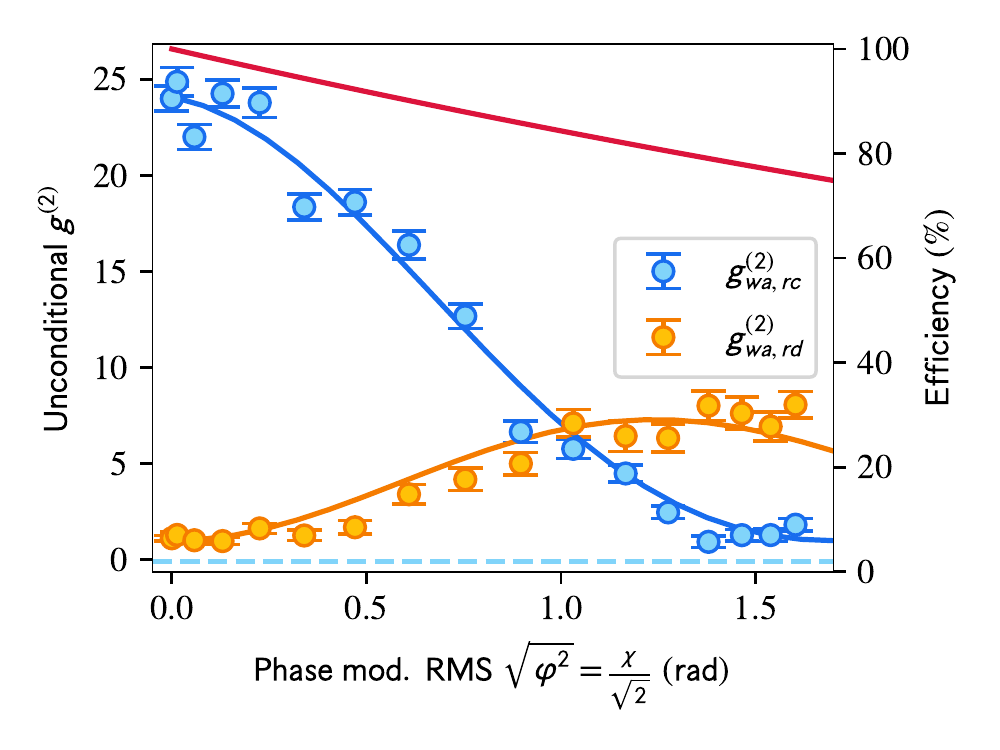}\caption{Beam-splitter operation at the single spin-wave level. Measured
		second order correlation between the $wa$ write photon mode for two
		read-out detection modes $rc$ ($k_{y}^{r}+k_{y}^{w}=0$) and $rd$ ($k_{y}^{r}+k_{y}^{w}=k_{g}$)
		as a function of phase modulation RMS $\sqrt{\langle\varphi^{2}\rangle}$ (left axis; dashed line corresponds to 0), and predicted total diffraction efficiency in all modes (red line, right axis). Curves correspond to the theoretical
		prediction with a the initial correlation and decay as fit parameters.\label{fig:bimspliter}}
\end{figure}

Here we evaluate the performance of spin-wave splitter in the few-mode
quantum interference experiment as well as give explicit expressions
used to analyze spin-wave diffraction presented in Fig. 2 of the main text. To explicitly
express the transformation of the spin-wave creation operator $\swd_{K_{y}}\xrightarrow{\varphi_{S}(y)}\swds_{K_{y}}$,
describing the result of imprinting on atoms a sine-shaped phase pattern
$\varphi_{S}(y)=\chi\sin(k_{g}y+\vartheta)$ we may use the well-known
Jacobi-Anger identity to expand the modulation term into more convenient
form: $\exp(i\chi\sin(k_{g}y+\vartheta))=\sum_{n=-\infty}^{\infty}J_{n}(\chi)\exp(in(k_{g}y+\vartheta))$,
where $J_{n}$ is the $n$-th Bessel function of the first kind. With
this expansion we easily obtain that: 
\begin{equation}
\swds_{K_{y}}=\sum_{n=-\infty}^{\infty}J_{n}(\chi)\exp(in\vartheta)\swd_{K_{y}+nk_{g}}.
\end{equation}
Using this formula with $\chi$ chosen so that $J_{0}(\chi)=J_{1}(\chi)$
and neglecting terms with $n>1$, we get the transformation used to
describe the HOM interference of spin waves (i.e. a 50:50 beamsplitter
transformation). Generally, this expression allows us also to predict
the unconditional $g^{(2)}$ function dependence on modulation RMS
amplitude. Choosing two modes separated by $\Delta K_{y}=k_{g}$ (i.e
$rc$ and $rd$) and neglecting the contribution of weak thermal state
split into the $rc$ mode (i.e. assuming modes $va$ and $vb$ reside
in vacuum) we can write: 
\begin{gather}
g_{wa,rc}^{(2)}=1+\alpha J_{0}(\chi)^{2}e^{-\gamma \chi},\\
g_{wa,rd}^{(2)}=1+\alpha J_{1}(\chi)^{2}e^{-\gamma \chi}. 
\end{gather}
As $\chi=\sqrt{2\langle\varphi^{2}\rangle}$. We heuristically include the exponential decay with a constant $\gamma$ to model dephasing due to deviations of the ac Stark grating from a perfect sine wave.
The results for the fit parameters are $\alpha=23.1\pm0.3$ and $\gamma=0.27\pm0.05\ \mathrm{rad}^{-1}$. The cross-correlation function without the modulation applied corresponds to $g_{wa,rc}^{(2)}\ (\mathrm{no\ modulation})=\alpha+1=24.1\pm0.3$. This allows us to estimate the total efficiency at $\sqrt{\langle\varphi^{2}\rangle} = 1\ \mathrm{rad}$ as 84\%. The fit results are presented in Fig. \ref{fig:bimspliter} along with the efficiency corresponding to $e^{-\gamma \chi}$. The measurements for this section used a significantly higher write-out photon detection rate of $\sim4\times10^{-2}$, thus the value of cross-correlation is smaller than in HOM dip measurement (yet still above the nonclassicality bound).

To generate asymmetric spin-wave patterns, we use different phase
modulation. The modified version includes a ``second-harmonic''
term:

\begin{equation}
\varphi_{S}(y)=\chi_{1}\sin(k_{g}y+\vartheta_{1})+\chi_{2}\sin(2k_{g}+\vartheta_{2}).
\end{equation}

By taking $\chi_{1}/\chi_{2}=2.5$, we observe that spin-wave diffraction
occurs predominantly in one direction (as in Figs. 2c and 2d). By
changing the relative phase between the two terms above, i.e. $\Delta\vartheta=\vartheta_{1}-\vartheta_{2}$,
we can steer the direction of diffraction. In particular, for Fig.
2(c) we selected $\Delta\vartheta=0$ and for Fig. 2(d) we set $\Delta\vartheta=\pi$. 

\subsection{Quantum character certification}
\begin{figure}
	\includegraphics[width=0.85\columnwidth]{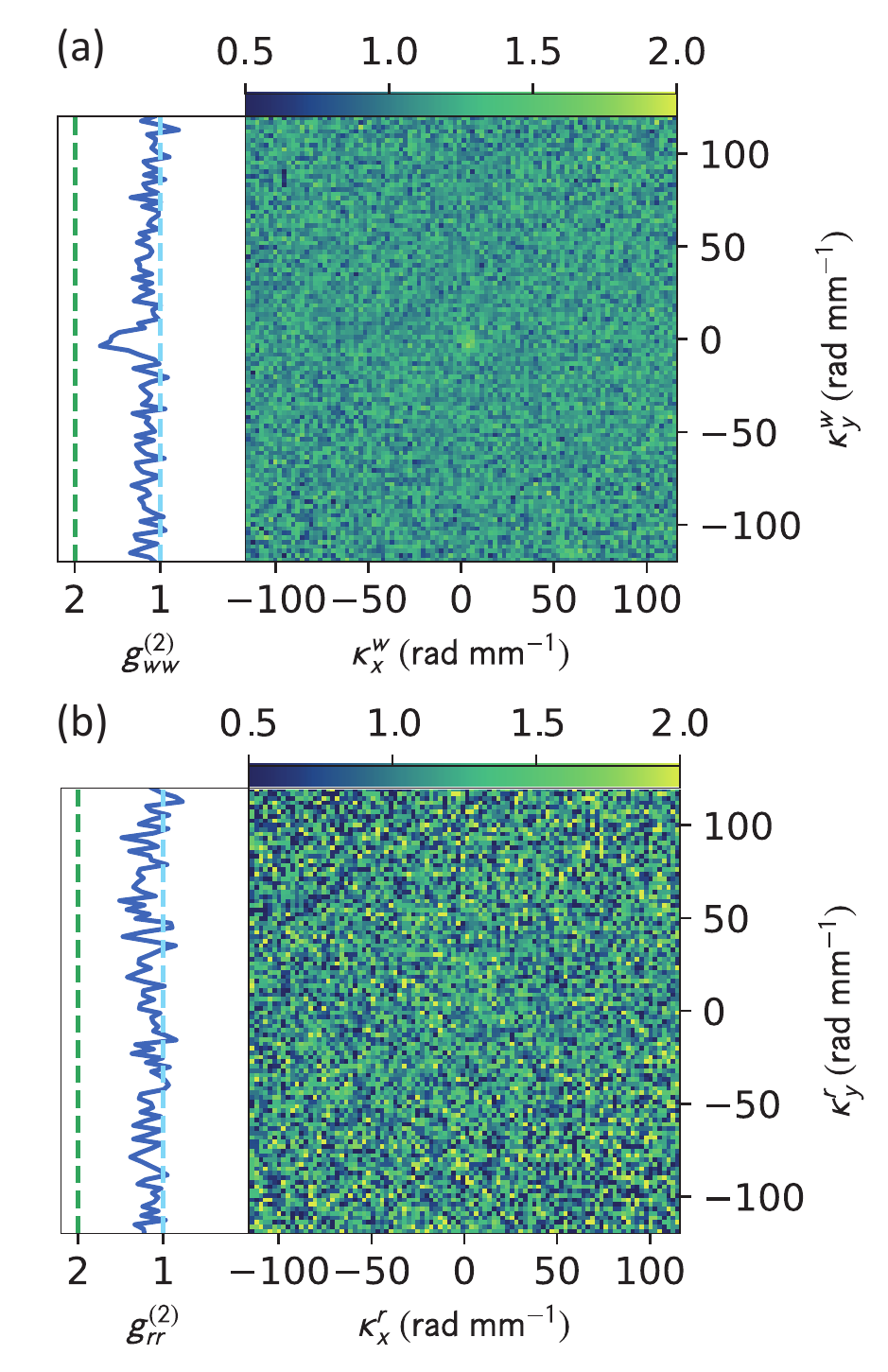}
	
	\caption{Autocorrelations for write-out and read-out photons. The autocorrelations
		in both cases were measured by splitting the relevant photon stream
		into two regions of the I-sCMOS detector (see Fig.
		\ref{fig:eksper}) and effectively performing a multimode HBT measurement. Due to a low photon number caused by detection inefficiency, we do
		not observe significant bunching for read-out photons [panel (b)],
		while for write-out photons a weak autocorrelation peak is observed [panel
		(a)]. The autocorrelation functions are expressed in terms
		of wavevector-difference variables i.e. $\kappa_{x}^{w}=k_{x}^{w1}-k_{x}^{w2}$,
		$\kappa_{y}^{w}=k_{y}^{w1}-k_{y}^{w2}$, $\kappa_{x}^{r}=k_{x}^{r1}-k_{x}^{r2}$
		and $\kappa_{y}^{r}=k_{y}^{r1}-k_{y}^{r2}$. Cross sections are taken
		around $\kappa_{x}^{w}=0$ and $\kappa_{x}^{r}=0$ for panel (a)
		and (b), respectievly.\label{fig:altokorelacje}}
	
\end{figure}
We use appropriate values of the second-order Glauber correlation
function to certify nonclassicality of photon-counting statistics
\cite{Paul1982}. For the wavevector-resolved measurements we utilize
multiplexing of many modes and evaluate the averaged correlation function
\begin{multline}
g_{rw}^{(2)}(k_{x}^{r}+k_{x}^{w},k_{y}^{r}+k_{y}^{w})=\\
\frac{\int\langle\hat{n}_{r}(k_{x}^{r},k_{y}^{r})\hat{n}_{w}(k_{x}^{w},k_{y}^{w})\rangle\mathrm{d}(k_{x}^{r}-k_{x}^{w})\mathrm{d}(k_{y}^{r}-k_{y}^{w})}{\int\langle\hat{n}_{r}(k_{x}^{r},k_{y}^{r})\rangle\langle\hat{n}_{w}(k_{x}^{w},k_{y}^{w})\rangle\mathrm{d}(k_{x}^{r}-k_{x}^{w})\mathrm{d}(k_{y}^{r}-k_{y}^{w})}
\end{multline}
Nonclassicality of the generated state between write-out and read-out photons
is certified by violating the Cauchy-Schwarz inequality: $[g_{rw}^{(2)}]^{2}\leq g_{rr}^{(2)}g_{ww}^{(2)}$.
Since we measured auto-correlation values $g_{rr}^{(2)}$, $g_{ww}^{(2)}<2$
(see Fig. \ref{fig:altokorelacje}), a conservative
bound on nonclassicality is also given by $g_{rw}^{(2)}>2$, which
we have used throughout the article. A high value of $g_{rw}^{(2)}$
is a good indication that single read-out photons (or spin waves) is characterized
by good purity (or in other words will be close to a single-excitation
Fock state). A more direct indicator is the conditional $g_{r1,r2|w}^{(2)}$,
which we found for the interference experiment presented in Fig. 4 of the main text
as $g_{rc,rc|wa}^{(2)}=0.67\pm0.04<1$ and $g_{rd,rd|wb}^{(2)}=0.69\pm0.04<1$
for maximum $\Delta K_{x}$ separation. Finally, the Hong-Ou-Mandel
interference is customarily witnessed by the observed depth (visibility
$\mathcal{V}$) of the dip in coincidences larger than $\mathcal{V}=0.5$.
Here we used a normalized conditional correlation function which without
interference equals 1 (we measured $0.93\pm0.44$ for the largest
separation of modes, which is consistent with the Gaussian mode shape).
We may thus estimate the visibility as $\mathcal{V}=1-g_{rc,rd|wa,wb}^{(2)}=0.80\pm0.06>0.5$,
certifying a non-classical character of interference. Only slightly
lower visibility of $\mathcal{{V}}=0.66\pm0.01>0.5$ is observed for
the interference of a single spin-wave with the weak thermal spin-wave
state. The value of conditional $g_{rc,rd|wa}^{(2)}<1$ for this case
also certifies single excitation character of the spin wave in mode
$ra$, obtained using the HBT measurement performed at the spin-wave
level (as opposed to using optical elements to implement the beam-splitter
at the photonic level). Importantly, this figure of merit better witnesses
single-excitation character than $g_{rc,rc|wa}^{(2)}$ or $g_{rd,rd|wb}^{(2)}$.
All given errors correspond to one standard deviation of the Poissonian
counts statistics.
\subsection{Modeling the beamsplitter network}
\begin{figure}
	\includegraphics[width=0.85\columnwidth]{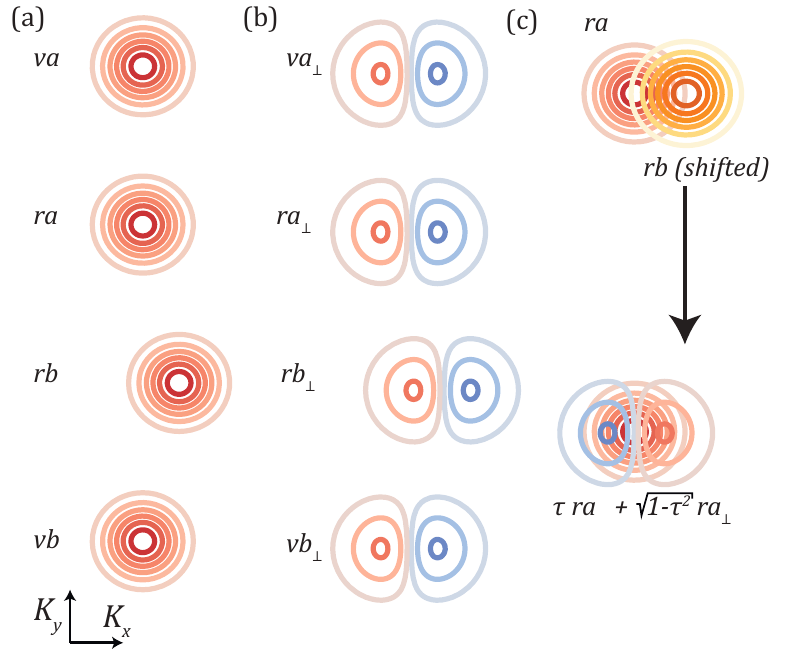}
	\caption{Choice of wavevector-space basis for the beamsplitter network theoretical analysis: (a) proper modes $ra$ and $rb$ are selected in positions of the collection fibers, (b) orthogonal modes $ra_\perp$ and $rb_\perp$ are selected to allow construction of a shifted proper mode, as demonstrated in (c) where $rb$ is constructed as a superposition of $ra_\perp$ and $ra$. The in top panel of (c) the $ra$ mode is shown for reference, while the bottom panel of (c) shows two modes into which $rb$ is decomposed. For small shifts $\Delta K_x$ the overlap $\tau\approx1$ and the orthogonal modes are simply first Hermite-Gauss modes.\label{fig:modyorto}}
\end{figure}
While in Fig. 5 of the main text the shapes of the presented theoretical curves roughly correspond to the shape of the fiber detection mode $u_\perp(\mathbf{K}_\perp)$ (in terms their widths and very roughly shapes), to properly describe these shapes as well as to obtain quantitative agreement of correlation functions a more elaborate model is required. We model the process in three steps: heralded generation, manipulation, read-out and detection that also adds noise during photons' detection. For the generation stage we assume a total of 8 pairs of squeezed modes: four of them correspond to read-out modes $ra$, $rb$, $va$ and $vb$ as denoted in Fig. 4 of the main text. They have associated write-out photon modes $wa$, $wb$ that we detect, while the other write-out modes are not detected. These modes are arranged in the $K_y$ direction. In the $K_x$ direction for each of these modes we also assume another pair of \textit{orthogonal} modes ($ra_\perp$, $rb_\perp$, etc.), that are squeezed in the same way (these constitute another 4 mode pairs, yielding a total of 8). As we change mode positions in the $K_x$ direction we assume that proper modes move accordingly (see Fig. \ref{fig:modyorto}).
During manipulation a three-way splitter operation is applied in the $K_y$ direction, mixing modes with same $K_x$. The corresponding three-way splitter Bogoliubov transformation for modes $ra$, $rb$, $va$ and $vb$ is:
\begin{multline}
\mathrm{Tws}(vb,rb,ra,va) =\\ \frac{1}{\sqrt{3}}\left(
\begin{array}{cccccccc}
1 & 0 & 1 & 0 & 0 & 0 & 0 & 0 \\
0 & 1 & 0 & 1 & 0 & 0 & 0 & 0 \\
-1 & 0 & 1 & 0 & 1 & 0 & 0 & 0 \\
0 & -1 & 0 & 1 & 0 & 1 & 0 & 0 \\
0 & 0 & -1 & 0 & 1 & 0 & 1 & 0 \\
0 & 0 & 0 & -1 & 0 & 1 & 0 & 1 \\
0 & 0 & 0 & 0 & -1 & 0 & 1 & 0 \\
0 & 0 & 0 & 0 & 0 & -1 & 0 & 1 \\
\end{array}
\right)
\end{multline}
where subsequent rows correspond to operators: $\sw_{vb},\ \swd_{vb},\ \sw_{rb},\ \swd_{rb},\ \sw_{ra},\ \swd_{ra},\ \sw_{va},\ \swd_{vb}$.
However, if the modes are separated in the $K_x$ direction we need to consider that for instance $ra$ mixes both with $rb$ and $rb_\perp$, as in the following beamsplitter transformation:
\begin{multline}
\mathrm{Bs}(rb,rb_\perp) =\\ \left(
\begin{array}{cccc}
\tau  & 0 & \sqrt{1-\tau ^2} & 0 \\
0 & \tau  & 0 & \sqrt{1-\tau ^2} \\
-\sqrt{1-\tau ^2} & 0 & \tau  & 0 \\
0 & -\sqrt{1-\tau ^2} & 0 & \tau  \\
\end{array}
\right)
\end{multline}
We thus apply the three-way splitter transformation sandwiched between a basis change transformation that mixes $rb$ and $rb_\perp$ (we may disregard mixing of unheralded modes, since they are not detected). The basis change is a beamsplitter transformation with transmission given by mode overlap equal $\tau=\int \mathrm{d}K_x \mathrm{d}K_y u_\perp(K_x,K_y) u^*_\perp(K_x+\Delta K_x,K_y+\Delta K_y)$ (in our case $\Delta K_y=0$). The results is a Gaussian function for our Gaussian-shaped modes. The total transformation, including initial two-mode squeezings, is:
\begin{multline}
\mathrm{Bs}(rb,rb_\perp)^T \\ \left(\mathrm{Tws}(vb,rb,ra,va) \otimes \mathrm{Tws}(vb_\perp,rb_\perp,ra_\perp,va_\perp)\right) \\ \mathrm{Bs}(rb,rb_\perp) \left(\bigotimes_{i=1}^8 \mathrm{Sq}(w_i,r_i) \right) 
\end{multline}
where the two-mode squeezing transformation is:
\begin{multline}
\mathrm{Sq}(w,r) =  \left(
\begin{array}{cccc}
\frac{1}{\sqrt{1-p^2}} & 0 & 0 & \frac{p}{\sqrt{1-p^2}} \\
0 & \frac{1}{\sqrt{1-p^2}} & \frac{p}{\sqrt{1-p^2}} & 0 \\
0 & \frac{p}{\sqrt{1-p^2}} & \frac{1}{\sqrt{1-p^2}} & 0 \\
\frac{p}{\sqrt{1-p^2}} & 0 & 0 & \frac{1}{\sqrt{1-p^2}} \\
\end{array}
\right)
\end{multline}
with $p$ being the fundamental pair generation probability, and subsequent rows correspond to operators $\sw_{r},\ \swd_{r},\ \sw_{w},\ \swd_{w}$.

Finally, we obtain output modes $rc$ and $rd$ and calculate proper expectation values on vacuum, including influence of dark counts as $p_\mathrm{dark}/\eta$ (where $\eta$ is the net detection efficiency) in the read-out modes (write-out modes contain several times more photons and thus the dark counts there are negligible). For instance, within this framework the correlation function $g^{(2)}_{rc,rd|wa,wb}$ is calculated as:
\begin{multline}
g^{(2)}_{rc,rd|wa,wb}=\\ (\langle \hat{n}_{rc} \hat{n}_{rd} \hat{n}_{wa} \hat{n}_{wb} \rangle + p_\mathrm{dark}/\eta \langle \hat{n}_{rc} \hat{n}_{wa} \hat{n}_{wb} \rangle +  \\ p_\mathrm{dark}/\eta \langle \hat{n}_{rd} \hat{n}_{wa} \hat{n}_{wb} \rangle +  (p_\mathrm{dark}/\eta)^2 \langle \hat{n}_{wa} \hat{n}_{wb} \rangle) \langle \hat{n}_{wa} \hat{n}_{wb} \rangle / \\ (\langle \hat{n}_{rc} \hat{n}_{wa} \hat{n}_{wb} \rangle+p_\mathrm{dark}/\eta \langle \hat{n}_{wa} \hat{n}_{wb} \rangle) \\ (\langle \hat{n}_{rd} \hat{n}_{wa} \hat{n}_{wb} \rangle+p_\mathrm{dark}/\eta \langle \hat{n}_{wa} \hat{n}_{wb} \rangle) = \\ = \eta ^2 \left(9 p^2+6 p \left(\tau ^2+1\right)+\left(\tau ^2-1\right)^2\right)-\\ 6 \eta  p_\mathrm{dark} \left(3 p^2+p \left(\tau ^2-2\right)-\tau ^2-1\right)+9 (p-1)^2 p_\mathrm{dark}^2 \\ / \left(\eta  \left(3 p+\tau ^2+1\right)-3 (p-1) p_\mathrm{dark}\right)^2.
\end{multline}
Analogous expressions are obtained for all other correlation functions using a computer algebra system. We obtain a very good agreement with experimental observations in all cases by setting $p=0.05$ and $p_\mathrm{dark}/\eta =0.017$. Note that to explain the slight drop of the $g^{(2)}_{wa,rc}$ in Fig. 5(c) we introduced a residual misalignment of modes $wa$ and $rc$ that scales as $0.29 \Delta K_x$ - the dashed curve account for that.
\subsection{Phase matching at readout}
\begin{figure*}
	\includegraphics[width=0.7\textheight]{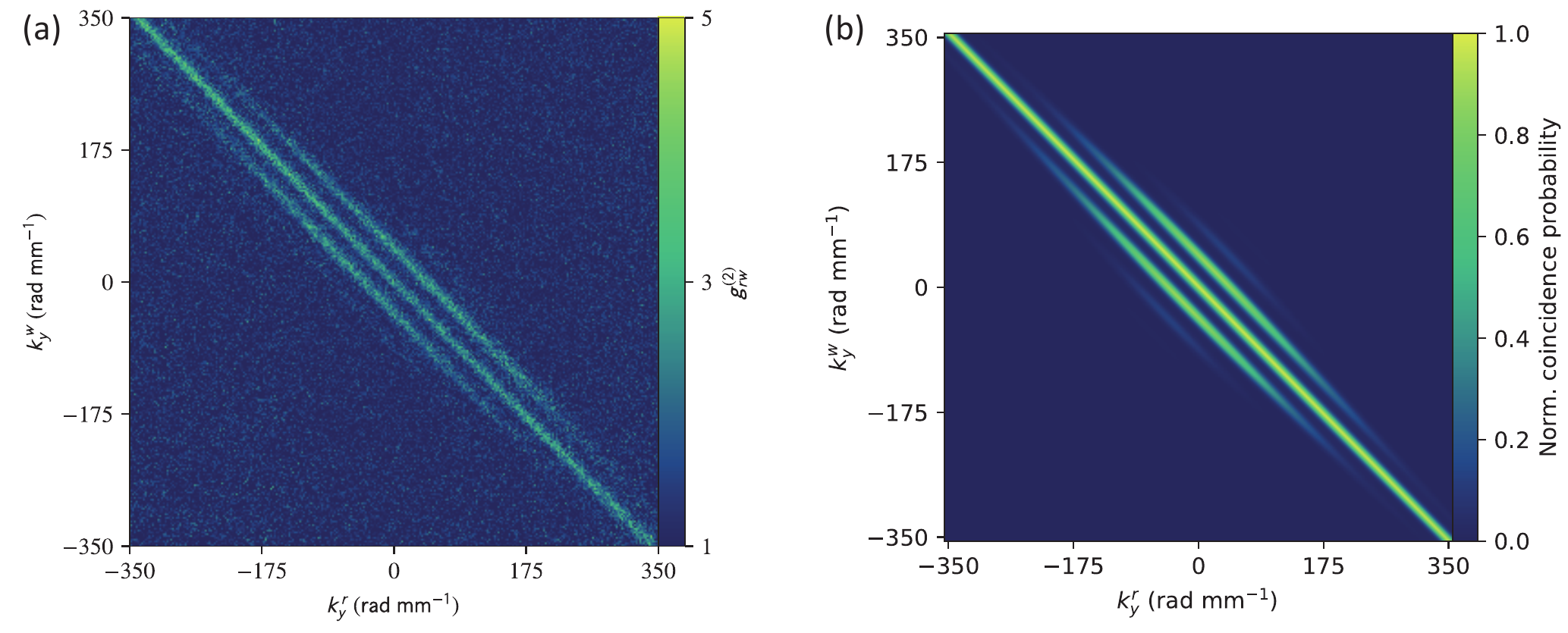}\caption{Influence of phase matching on the readout of diffracted spin-waves.
		Panel (a) portrays normalized write-out--read-out coincidences in the
		form of the second order correlation function (obtained from the same
		data as Fig. 3b in the main text). We compare this experimental
		result with the calculated normalized readout efficiency of the reshaped
		spin-waves presented in panel (b) [corresponding to the same
		write photons as in (a)]. Strong influence of phase-matching
		is evident, as read-out is only efficient for values of $k_{y}^{r}$
		around 0.\label{fig:dopasowankofazkowe}}
\end{figure*}
To estimate the readout efficiency we resort to a set of equations
describing a classical, optical $\mathcal{E}_{r}(\mathbf{k}_{\perp},z,t)$ (in terms of the slowly varying envelope)
and spin-wave field $\rho_{hg}(\mathbf{K}_{\perp},z,t)$ (atomic coherence)
at the readout stage. We assume a uniform intensity profile for the
Raman pump and include the atom number density as $S(\mathbf{r}_{\perp},z,t)=\sqrt{\mathscr{N}(\mathbf{r}_{\perp},z)}\rho_{hg}(\mathbf{r}_{\perp},z,t)$,
where $\rho_{hg}(\mathbf{r}_{\perp},z,t)$ is a coherence averaged overs atoms in a small
volume around $\mathbf{r}=(\mathbf{{r}}_{\perp},z)=(x,y,z)$ and $\mathscr{N}(\mathbf{r}_{\perp},z)=\mathscr{N}_{0}\exp(-|\mathbf{r}_{\perp}|^{2}/\sigma_{\perp}^{2}-z^{2}/\sigma_{z}^{2})$. The total number of atoms is $N=\int \mathrm{d}\mathbf{r} \mathcal{N}(\mathbf{r})$. After quantization and in the wavevector domain $\hat{S}(\mathbf{K})=(2\pi)^{-3/2} \sum_n^N \mathcal{N}(\mathbf{r}_n)^{-1/2} e^{i \mathbf{K}\cdot\mathbf{r}_n} |h_n\rangle\langle g_n|$ corresponds to a continuous spin-wave density operator, and in general the discrete operators that we use in the main text correspond to $\hat{S}_u = \int \mathrm{d}\mathbf{K} \hat{S}(\mathbf{K}) \tilde{u}^*(\mathbf{K})$, where $\tilde{u}(\mathbf{K})$ is a normalized mode function. For operators that we denote $\sw_\mathbf{K}$ we have $u_\mathbf{K}(\mathbf{r})=\sqrt{\mathcal{N}(\mathbf{r})/N} e^{i \mathbf{K}\cdot\mathbf{r}}$ with $u$ and $\tilde{u}$ being related via the Fourier transform.

Importantly, we include the diffraction term within the wide-angle,
slowly-varying envelope approximation \cite{Hadley1992,Trippenbach2002,Leszczynski2017} derived from  UPPE (unidirectional pulse propagation equation) \cite{Couairon2011}
and write the equations in the frame of reference co-moving with the
readout pulse:

\begin{multline}
\frac{{\partial\mathcal{\mathcal{{\mathcal{{E}}}}}}_{r}(\mathbf{{k}}_{r\perp},z,t)}{\partial z}- i\left(\sqrt{k_{r}^{2}-
	\mathbf{k_{\perp}}^{2}}-k_{r}\right)\mathcal{{E}}_{r}(\mathbf{{k}}_{r\perp},z,t)
=\\-\frac{i k_r}{\epsilon_0 \hbar} d_{ge} \frac{d_{he} \mathcal{E}_R}{2\Delta}e^{i(k_r-k_R)} \mathscr{F}_\perp(\sqrt{\mathcal{N}})*S(\mathbf{{K}}_{\perp},z,t)
\end{multline}

\begin{equation}
\frac{{\partial S}(\mathbf{{K}_{\perp}},z,t)}{\partial t}=-\frac{d_{he} \mathcal{E}_R^*}{4i\Delta} e^{i(k_R-k_r)} \mathscr{F}_\perp(\sqrt{\mathcal{N}})* \mathcal{E}_r(\mathbf{{k}}_{r\perp},z,t)
\end{equation}

with $\mathbf{k}_{r\perp}=\mathbf{k}_{R\perp}+\mathbf{K}_{\perp}$, $k_r$ and $k_R$ are length of read-out photon and read laser wavevectors, $d_{ij}$ are dipole moments of respective transitions, $\Delta$ is the detuning from the single-photon resonance, $\mathcal{E}_R$ field amplitude of the read laser, and $*$ and $\mathscr{F}_\perp$ denote convolution and the Fourier transform in the transverse coordinates, respectievely. Phase matching arising due to
oscillations caused by the $i\left(\sqrt{k_r^{2}-\mathbf{k_{\perp}}^{2}}-k_{r}\right)\mathcal{{E}}_{r}(\mathbf{{k}}_{r\perp},z,t)$
term emerges as the most essential factor influencing readout of reshaped
spin-waves. To quantify this effect it is enough to consider the first-order
approximation of the well-studied solution for the above set of equations,
for the case of null $\mathcal{E}_{r}$ field at the input (see Refs. \cite{Farrera2016,Raymer2004,Koodynski2012} for more complete ways of treatment).
Then, we may equivalently write that ${\partial_{t}S}(\mathbf{{k}_{\perp}},z,t)\approx0$
and easily solve the first equation as a Gaussian integral. By considering
the readout efficiency of a plane-wave spin-wave reshaped with a sine
grating with $k_{g}$, we arrive at very good agreement between measured
coincidence map presented in  Fig. \sref{fig:dopasowankofazkowe}{a}
and the theoretical prediction [Fig. \sref{fig:dopasowankofazkowe}{b}]
for phase-matching efficiency (here normalized to unity), indicating
that the phase-matching is the most essential wavevector-dependent
factor to the net readout efficiency. The main observed effect is
the fact that only spin waves with small $K_{y}$ can be read-out
efficiently after modulation. Remaining spin-wave are not lost, but
can be retrieved through manipulating their $K_{z}$ wavevector components
to restore the phase-matching. The problem can thus be alleviated
through use of the ac Stark gradient in the $z$-direction or rotation of the grating in the $y$-$z$ plane. Furthermore,
it is worth noting that the region for which the readout is naturally
efficient encompasses hundreds of usable modes (in terms of Schmidt
decomposition).

Numerical values of the parameters used for the calculation are $k_{g}=44\ \mathrm{rad\ mm}^{-1}$,
$\sigma_{\perp}=0.3\ \mathrm{mm}$, $\sigma_{z}=4\ \mathrm{mm}$,
$k_{r}=7899\ \mathrm{rad\ mm}^{-1}$. Correspondence between the write
photon wavevector and the spin-wave wavevector is calculated as $\mathbf{K}=\mathbf{k}^{W}-\mathbf{k}^{w}$.

\end{document}